\documentclass[sigconf]{acmart}
\usepackage{algorithmic}
\usepackage[ruled, linesnumbered]{algorithm2e}
\newlength\mylen

\usepackage{diagbox}
\usepackage{multirow}
\AtBeginDocument{%
  \providecommand\BibTeX{{%
    \normalfont B\kern-0.5em{\scshape i\kern-0.25em b}\kern-0.8em\TeX}}}
\usepackage{array}
\usepackage{subfigure}
\usepackage{balance}

\copyrightyear{2024}
\acmYear{2024}
\setcopyright{acmlicensed}\acmConference[WSDM '24]{Proceedings of the 17th ACM International Conference on Web Search and Data Mining}{March 4--8, 2024}{Merida, Mexico}
\acmBooktitle{Proceedings of the 17th ACM International Conference on Web Search and Data Mining (WSDM '24), March 4--8, 2024, Merida, Mexico}
\acmPrice{15.00}
\acmDOI{10.1145/3616855.3635830}
\acmISBN{979-8-4007-0371-3/24/03}

\begin{document}

\title{User Consented Federated Recommender System Against Personalized Attribute Inference Attack}


\author{Qi Hu}
\affiliation{%
  \institution{Department of CSE, HKUST}
  \city{Hong Kong SAR}
  \country{China}}
\email{qhuaf@connect.ust.hk}

\author{Yangqiu Song}
\affiliation{%
  \institution{Department of CSE, HKUST}
  \city{Hong Kong SAR}
  \country{China}}
\email{yqsong@cse.ust.hk}

\renewcommand{\shortauthors}{Qi Hu \& Yangqiu Song}

\begin{abstract}
Recommender systems can be privacy-sensitive. To protect users’ private historical interactions, federated learning has been proposed in distributed learning for user representations. Using federated recommender (FedRec) systems, users can train a shared recommendation model on local devices and prevent raw data transmissions and collections. However, the recommendation model learned by a common FedRec may still be vulnerable to private information leakage risks, particularly attribute inference attacks, which means that the attacker can easily infer users’ personal attributes from the learned model. Additionally, traditional FedRecs seldom consider the diverse privacy preference of users, leading to difficulties in balancing the recommendation utility and privacy preservation. Consequently, FedRecs may suffer from unnecessary recommendation performance loss due to over-protection and private information leakage simultaneously. In this work, we propose a novel \textit{user-consented federated recommendation system} (UC-FedRec) to flexibly satisfy the different privacy needs of users by paying a minimum recommendation accuracy price. UC-FedRec allows users to self-define their privacy preferences to meet various demands and makes recommendations with user consent. Experiments conducted on different real-world datasets demonstrate that our framework is more efficient and flexible compared to baselines. Our code is available at \url{https://github.com/HKUST-KnowComp/UC-FedRec}.
\end{abstract}

% In traditional FedRec, thee gradients are transferred between clients and servers for central model aggregating and local model updating and the process are often protected by differential privacy to avoid privacy leakage.

% %%
% %% The code below is generated by the tool at http://dl.acm.org/ccs.cfm.
% %% Please copy and paste the code instead of the example below.
% %%


% %%
% %% Keywords. The author(s) should pick words that accurately describe
% %% the work being presented. Separate the keywords with commas.

%   \includegraphics[width=\textwidth]{figure/sampleteaser.pdf}

\maketitle

\section{Introduction}

Recommender systems have gained considerable popularity in predicting users' interests in online services, such as e-commerce and social media \cite{resnick1997recommender}. 
Recently, deep learning based recommendation algorithms, which use interactions of users and items, and/or attributes with a parameterized network to predict users' preferences, have proven to be effective \cite{he2017neural,zhang2019deep}. However, the historical user and item interaction data and some users' attribute data are highly privacy-sensitive. With the growing attention to privacy preservation and the changes in privacy regulations such as the General Data Protection Regulation (GDPR), balancing privacy protection and recommendation accuracy is becoming increasingly critical \cite{zhang2014privacy, shin2018privacy}.
\begin{figure}[t] 
  \centering
  \includegraphics[width=0.9\linewidth]{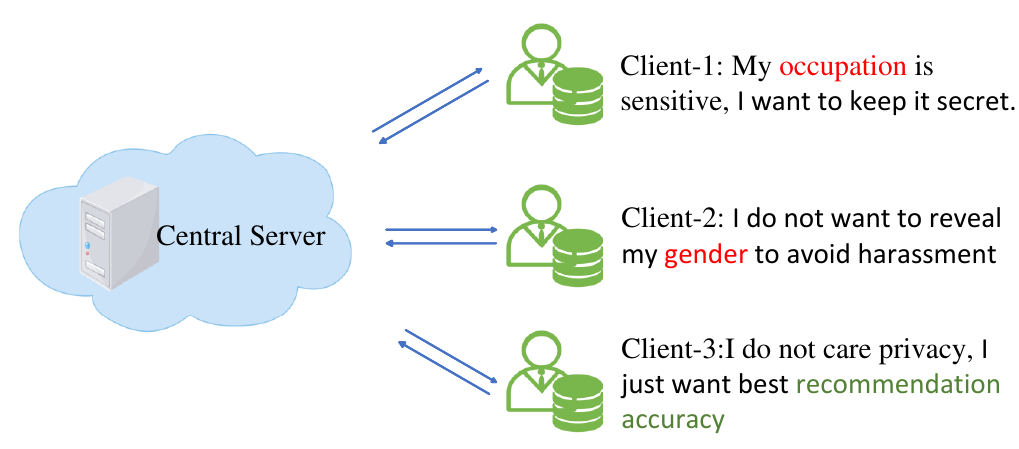}
  \caption{Overview of the user consented FedRec framework. Different users have various privacy or utility preferences.}
  \label{Fig:overview}
\vspace{-0.4cm}
\end{figure}

To protect users' raw data, federated learning has been proposed, which only exchanges the gradients between users and the server \cite{konevcny2016federated,mcmahan2017communication}. Based on federated learning, federated recommender (FedRec) systems have been developed to address privacy issues by decentralizing the training process. In such systems, raw user-item interaction data is decoupled from the model training \cite{muhammad2020fedfast, cormode2018privacy}. Despite avoiding raw data collection and transmission, several issues still need to be addressed with FedRec systems. First, the personalization of users learned by FedRec, based on their past behaviors, poses the risk of user-level private information leakage \cite{ateniese2015hacking,gong2016you,zhang2021membership}. Research has shown that a high-quality personalized federated learning model is vulnerable to attribute inference attacks to reveal participants' personal information \cite{wang2019beyond, calandrino2011you}. We use attackers to predict users' attributes from some FedRecs \cite{minto2021stronger, muhammad2020fedfast} trained on MovieLens \cite{harper2015movielens}, the results in Table \ref{tab:intro} show that those FedRecs face attribute inference attacks. Second, there is a tradeoff between privacy protection and recommendation utility. However, privacy demands differ among participants, and existing FedRec systems provide the same privacy protection for all users without considering their specific demands. This results in users with strong privacy demands are not satisfied, whereas users with weak privacy demands pay an unnecessary recommendation performance price. Third, every user's privacy demands are not static, and traditional FedRec systems mainly focus on privacy protection in the training stage, making it difficult for users to modify privacy settings \cite{minto2021stronger, perifanis2022federated}. Consequently, the balance between utility and privacy is not flexible for users in traditional FedRec systems.  

To meet the personalized privacy demands of various users, we propose a new framework called \underline{U}ser-\underline{C}onsented \underline{Fed}erated \underline{Rec}ommender System (UC-FedRec), where users can flexibly safeguard their sensitive personal information and decide on the tradeoff between individual privacy and recommendation accuracy. The UC-FedRec framework aims to satisfy users' diverse privacy demands while paying minimum recommendation accuracy costs. As illustrated in Figure \ref{Fig:overview}, clients have various privacy demands and utility preferences. Thus, the framework has a dual objective: 1) to meet privacy needs and protect private clients from inference attacks, and 2) to avoid other clients paying the unnecessary cost and retaining high-quality recommendation results. To achieve this goal, we leverage the different privacy preferences of participants to train a set of sensitive attribute information filters. These filters can be optionally applied to the local recommendation models on client sides according to their privacy demands during the local training stage, so the model can flexibly eliminate the sensitive attribute information that users care about.

\begin{table}[]
\caption{Attribute inference attacks on FedRecs}
\label{tab:intro}
\small
\begin{tabular}{cccccc}
\hline
\multicolumn{1}{c|}{\multirow{2}{*}{FedRecs}} & \multicolumn{2}{c|}{Utility}        & \multicolumn{3}{c}{Privacy} \\ \cline{2-6} 
\multicolumn{1}{c|}{}                         & NDCG  & \multicolumn{1}{c|}{Recall} & Gender & Age   & Occupation \\ \hline 
\multicolumn{1}{c|}{LDP-Rec}                  & 0.694 & \multicolumn{1}{c|}{0.433}  & 0.685  & 0.296 & 0.136      \\
\multicolumn{1}{c|}{Fedfast-BPR}              & 0.72  & \multicolumn{1}{c|}{0.46}   & 0.745  & 0.353 & 0.15       \\ \hline
\multicolumn{3}{c}{Random Attacker}                                                 & 0.5    & 0.141 & 0.05       \\ \hline
\end{tabular}
\vspace{-0.4cm}
\end{table}
 
We summarize our main contributions as follows:
\begin{itemize}
    \item To the best of our knowledge, our work is the first user-consented FedRec framework. By applying sensitive attribute filters, users can flexibly protect their personal sensitive private information with proper configuration.
    \item We introduce an adversarial framework in FedRec and propose a personalized privacy-aware algorithm to meet users' diverse privacy and utility preferences. Our framework can meet the different privacy needs of users while minimizing the recommendation accuracy cost.
    \item We quantify the privacy leakage problem in FedRec and evaluate the proposed framework in privacy protection and utility on two real-world datasets. The results indicate that the framework can flexibly deal with various and changing user privacy preferences. 
\end{itemize}

\section{Related Work \label{sec:related}}
% \begin{table}
%   \caption{Notations and explanations.}
%   \label{tab:notation}
% % \begin{tabular}{@{}c p{5cm}<{\centering}}
% \begin{tabular}{@{}cc@{}}
% \toprule
% \multicolumn{1}{c|}{\textbf{Notation}} & \textbf{Explanation}                                                                                   \\ \midrule
% \multicolumn{1}{c|}{$R$}                & Recommender system                                \\ \hline

% \multicolumn{1}{c|}{$\mathcal{U}$}      & \begin{tabular}[c]{@{}c@{}}User set of $R$\\ with each user $u$, where $u \in \mathcal{U}$\end{tabular} 
% \\ \hline

% \multicolumn{1}{c|}{$\mathcal{I}$}      & \begin{tabular}[c]{@{}c@{}}Item set of $R$\\ With each item $i$, where $i \in \mathcal{I}$\end{tabular} 
% \\ \hline

% \multicolumn{1}{c|}{$f_\theta$, $s_\psi$}          & \multicolumn{1}{c}{Representation and score function in $R$} \\ \hline                               

% \multicolumn{1}{c|}{$e$, $h$}          & \multicolumn{1}{c}{Embedding and final representation of user/item}
%  \\ \hline 

% \multicolumn{1}{c|}{$\mathcal{T}_u$}    & Private attributes set for $u$  
% \\ \hline

% \multicolumn{1}{c|}{$g_{\phi_t}$}          & \multicolumn{1}{c}{Attribute filter for attribute $t$}    
% \\ \hline

% \multicolumn{1}{c|}{$\beta$, $\lambda$, $\delta$}    & Privacy related parameters 

% \\ \bottomrule
% \end{tabular}
% \end{table}
In this section, we briefly summarize the related work. Our work is closely related to collaborative filtering and privacy-preserving systems.
\subsection{Collaborative Filtering}
% \noindent {\bf Collaborative Filtering.}
Collaborative filtering (CF) is one of the most popular approaches in recommender systems and has been widely used in real-world systems. Having the assumption that people with similar historical interactions will have similar preferences, CF models, such as those proposed in \cite{ebesu2018collaborative, he2017neural, berg2017graph} parameterize users and items and their interactions by vectorized representations. Based on the representations, CF models predict the interactions by vector computations like inner product \cite{koren2009matrix}. To improve the recommendation accuracy and solve the cold-start problem, existing works focus on improving the quality of the representations. For example, some studies made use of side information such as user/item relations \cite{fan2019graph, xin2019relational} and external knowledge graph \cite{wang2019kgat}. Some works \cite{he2017neural,xue2017deep} utilized deep learning models to extract interaction features in user-item interactions. Some GNN-based collaborative filtering techniques \cite{wang2019neural, berg2017graph, chen2020revisiting} were proposed to exploit the high-order connectivity from user-item interactions. Most collaborative filtering techniques are centralized, requiring users to share their private historical interactions with the server. With the increasing concern about the privacy problem, decentralized collaborative models \cite{ammad2019federated, chai2020secure} were proposed to avoid sensitive data sharing. However, the models learned by the recommender system are personalized and are unbalanced for sensitive variables \cite{chen2020bias, chouldechova2017fair, bose2019compositional} and federated learning faces the problem same as the central system.

% \noindent {\bf Privacy preserving.}
\subsection{Privacy-Preserving Systems}
To overcome the privacy leakage problem, various frameworks are proposed. One direction targets the private information in learned models. For example, some studies use perturbation and differential privacy to prevent an adversary from inferring a targeted user's private information \cite{balu2016differentially, luo2014privacy, zhang2021graph, jia2018attriguard}. Some works propose to use adversarial learning to protect users' private personal attributes \cite{beigi2020privacy, he2020mining}. Some propose to disentangle the private information from utility tasks \cite{meng2018personalized, hu2022learning, hu2023independent}. However, these methods provide privacy preservation in central learning and cannot be applied in distributed learning where personal data is kept on local devices. To overcome the central data collection problem, federated learning was proposed, allowing data owners to collaboratively build a shared privacy-preserving decentralized model in distributed. \cite{mcmahan2017communication, konevcny2016federated, yang2019federated}. 
Many works are proposed to learn federated recommender systems with various privacy guarantees \cite{ammad2019federated, wu2021fedgnn, muhammad2020fedfast}.
Differential privacy (DP) \cite{wu2021fedgnn,  cormode2018privacy} is commonly applied in the context of distributed computing. It adds random noise to true data records such that two arbitrary records have close probabilities to generate the same noisy data record. It provides data anonymous that will not reveal individual information in sensitive information collection and analysis \cite{ren2018textsf}. However, DP in federated learning is designed for the model weights or update of information transmission and is not suitable for the problem of inference attacks and privacy leakage in the learned federated recommender model.

\section{User Consented Federated Recommendation \label{sec:problem_sec}}

In this section, we present the user-consented FedRec. It meets users' privacy demands by utilizing different privacy preferences to eliminate specific sensitive information accordingly.

\subsection{Preliminary \label{sec:pre}}

Following general settings of recommender systems \cite{kabbur2013fism, he2017neural, wang2019neural}, we denote a recommender system $R$ containing a set of users and items represented by $\mathcal{U} = \{u_1, u_2, \cdots, u_{|\mathcal{U}|}\}$ and $\mathcal{I} = \{i_1, i_2, \cdots, i_{|\mathcal{I}|}\}$ respectively. Denote the interaction matrix between users and items as $\mathbf{Y} \in \{0, 1\}^{|\mathcal{U}|\times |\mathcal{I}|}$, where the value of $1$ indicates that there is an interaction between corresponding user $u$ and item $i$ while the value $0$ means that user $u$ is not interested in the item $i$ or user is not aware of the existence of item $i$. Due to privacy concerns, federated learning is adopted. Users' data is stored in local devices to protect privacy, each user $u$ holds its own historical interaction records $\mathbf{Y}_u$. The recommendation model can be summarized as estimating user $u$'s preference on any item $i$ by learned latent user representations $h_u = f_\theta (u) \in \mathbb{R}^d$ and item representations $h_i = f_\theta (i) \in \mathbb{R}^d$, where $d$ denotes the representation size, so that:
\begin{equation}
  \hat{y}_{u,i} = s_\psi(h_u, h_i),
\end{equation}
where the scoring function $s_\psi(\cdot)$ can be dot product, multi-layer perceptions, etc., and $\hat{y}_{u,t}$ denotes the preference score for user $u$ on unobserved item $i$ which is usually presented in probability \cite{wu2020graph}. Representation function $f_\theta$ commonly has two parts: embedding layer which maps user/item to vectors $e_u$/$e_i$ and propagation layer which catches collaborative signals. The recommendation objective can be formalized as:
\begin{equation}
    \label{eqa:loss_rec}
    \min_{\theta, \psi}  \sum_{u \in \mathcal{U}, i \in \mathcal{I}} \mathcal{L} (s_\psi(f_\theta (u), f_\theta (i)), \mathbf{Y}),
\end{equation}
% \yqc{A loss function should be defined on a set of users, not an individual one. $\mathbf{Y}$ seems to be correct, but you need to indicate L is applied to all values in $\mathbf{Y}$.}
where $\mathcal{L}$ can be commonly used loss functions in recommender systems (e.g., BPR loss for implicit feedback  \cite{wang2019neural}). To solve the cold-start problem and improve recommendation accuracy, it is common for recommender systems to investigate a set of user attributes (e.g., occupation, location) $\mathcal{T}$. However, privacy can be a different definition for users \cite{raab1998distribution}. Though the information is kept on local devices, some users may find some attributes are sensitive and want to eliminate the information in FedRec. We denote user $u$'s personal sensitive attributes as $\mathcal{T}_u \subseteq \mathcal{T}$. 
% The user is denoted as $|\mathcal{T}|$-private users according to the number of sensitive attributes.

\subsection{Problem Definition \label{sec:problem}}

Given a FedRec where users have different privacy preferences, each user $u$ has its own private attribute set $\mathcal{T}_u$, and non-sensitive attributes $\mathcal{T}\backslash\mathcal{T}_u$ are revealed but are kept in local devices. The challenge is that if partial participants' attributes are leaked, the attacker can easily infer the unknown sensitive attributes that private users prefer not to reveal from the FedRecs. This is mainly because collaborative filtering models can catch the signals from similar behavior users. UC-FedRec aims to eliminate the information of sensitive attributes in the training stage with minimum recommendation utility cost so that the sensitive attribute leakage risks will be reduced. 
\begin{figure*}[ht]
  \centering
  \includegraphics[width=0.88\textwidth]{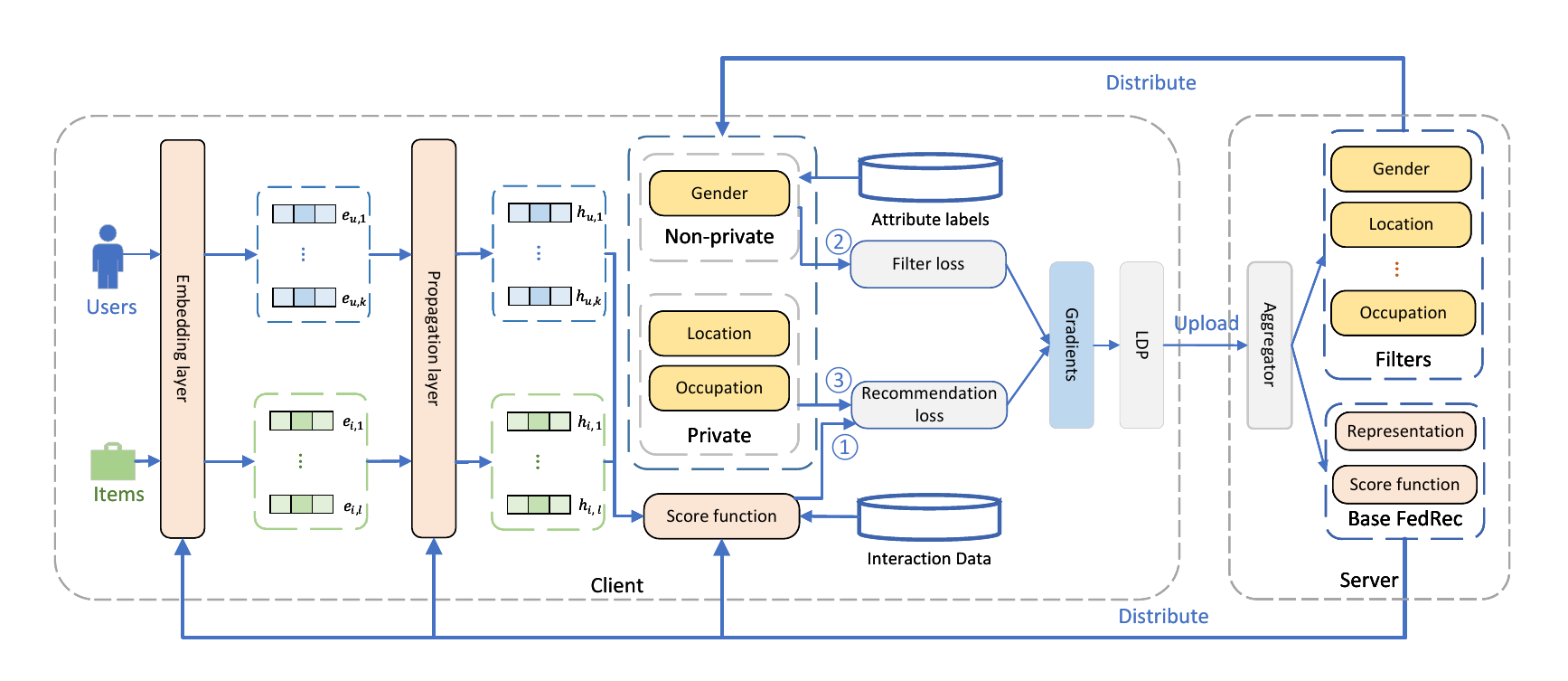}
  \caption{The overall architecture of UC-FedRec. The clients' training has three learning targets: \textcircled{1} to train the recommendation part as the primary goal; \textcircled{2} to train attribute distribution estimator on personal non-private attributes; \textcircled{3} to utilize estimators to eliminate private information in the representation model. Finally, it uploads the gradients to the server. The server aggregate and distribute the gradients after receiving local clients' upload.}
  \label{Fig:framework}
\end{figure*}
\subsection{Basic Framework  \label{sec:framework}}
Next, we introduce the basic structure of our proposed framework. UC-FedRec is a supplement of common FedRecs and uses arbitrary embedding-based FedRec as the base model. Compared to traditional FedRec, the critical module of UC-FedRec is the compositional sensitive attribute filters (distribution estimators). It can leverage the different privacy needs to learn sensitive attribute filters to protect private information from inference attacks. As shown in Figure \ref{Fig:framework}, it mainly consists of a central server and a set of user clients. On the client side, the filters are flexibly applied or trained by all the users according to their privacy preferences. User $u$’s privacy is protected by the $\mathcal{T}_u$ feature filters. These filters can eliminate the sensitive information in the representations. When the user regards some attributes as non-private, it takes the responsibility of training $\mathcal{T}\backslash\mathcal{T}_u$ feature distribution estimators. Each user learns user/item embeddings from its interaction data and feature extractors from its non-private attributes, and uploads filtered gradients of embeddings and non-private feature extractors. On the server side, the central server is responsible for aggregating the gradients and distributing the updated global models to clients. 

\subsection{Compositional Privacy Protection \label{sec:obj}}

\subsubsection{\textbf{Non-private attributes}}
In practice, FedRec faces the problem of sparsity and cold-start \cite{zhang2019deep}. It is common for FedRec to utilize user personal information to improve recommendation quality. Some users are willing to choose a part of non-private attributes to reveal for a better experience. Suppose user $u$ reals $\mathcal{T}\backslash\mathcal{T}_u$ features. We define the attribute probability distribution when given user $u$'s representations as $p(y_{u,t}|h_u)$, where $y_{u,t}$ is the label of attribute $t$ for user $u$. 
% \yqc{Shouldn't this be $y_{u,t}$? And I don't this we should model this joint probability. In your equation 3, you are modeling the prediction probability.} 
As the real distribution is unknown, and unfortunately the computation is expensive and in most cases intractable, it is necessary to introduce an auxiliary distribution $q_{\phi_t}(y_{u,t}|h_u)$ to approximate the posterior distribution. Users are responsible for the training of non-private attribute distribution estimators. We aim to have the $q_{\phi_t}(y_{u,t}|h_u)$ as close as possible to $p(y_{u,t}|h_u)$. Therefore, we use the Kullback–Leibler divergence (KL divergence) to measure the difference between two distributions \cite{kullback1951information} and minimize the distance:
% \yqc{This is not clear. Why do we need two distributions? Why can KL-D be used? Although you cite a reference here, you still need to introduce the motivation.}

\begin{equation}
\begin{split}
    &\min_{\theta, \phi_t} KL(p(y_{u,t}|h_u)\|q_{\phi_t}(y_{u,t}|h_u)) \\ 
  =& \min_{\theta, \phi_t} \mathbb{E}_{p(y_{u,t}|h_u)} \log p(y_{u,t}|h_u)  - \mathbb{E}_{p(y_{u,t}|h_u)} \log q_{\phi_t}(y_{u,t}|h_u)    \\
    \Leftrightarrow& \min_{\theta, \phi_{t}} -\mathbb{E}_{p(h_u, y_{u,t})}[\log q_{\phi_t}(y_{u,t}|h_u)] \quad \forall  t \in \mathcal{T}\backslash \mathcal{T}_u \text{.}
\end{split}
\end{equation}
To solve the objective function, we parameterize the $q_{\phi_{t}}$ function using a node classifier $g_{\phi_t}$ defined on the user representations. Suppose a set of users $\mathcal{U}_t$ consider $t$ as non-private attributes, then the objective function can be estimated as:
\begin{equation}
    \label{eqa:loss_non}
    \begin{split}
        \min_{\theta, \phi_t} \sum_{u \in \mathcal{U}_t} CE(g_{\phi_t}(f_\theta (u)), y_{u,t}),
    \end{split}
\end{equation}
where $CE(\cdot)$ is the cross entropy loss.
\subsubsection{\textbf{Private attributes}} The FedRec is at risk of leaking privacy, the fundamental reason is that the representations generated by the FedRec contain plenty of private information. Therefore, our goal is to eliminate the sensitive information in FedRec.  The sensitive attributes are expected to be independent from the learned user $u$'s representations, that is minimizing the mutual information $I(h_u;t)$ 
% \yqc{Again, it should be $y_{u, t}$. And here, instead of directly using $h_u$, you should introduce $g_{\phi_t}(f_\theta (u))$, as $h_u$ is not a distribution.} 
learned by FedRec, generalized to multiple attributes:
\begin{equation}
    \label{func:min_information}
    \min_\theta \sum_{t \in \mathcal{T}_u} I(h_u; t).
\end{equation}
To solve the Equation (\ref{func:min_information}), we leverage the upper bound $I_{v\text{CLUB}}(h_u;t)$ proposed in \cite{cheng2020club} and the minimizing attribute mutual information problem is equivalent to minimize the least upper bound:
\begin{equation}
\label{eq:filter_1}
\begin{split}
&\min_\theta \sum_{t \in \mathcal{T}_u} I(h_u; t)\\
\Leftrightarrow& \min_\theta \sum_{t \in \mathcal{T}_u} I_{v\text{CLUB}}(h_u;t) \text{.}
\end{split}
\end{equation}
To solve the Equation (\ref{eq:filter_1}), we utilize a objective function proposed in \cite{wang2021privacy}:
\begin{equation}
\label{eq:filter}
\begin{split}
\min_\theta \sum_{t \in \mathcal{T}_u} \max_{\phi_t} \mathbb{E}_{p(h_u, t)}[\log q_{\phi_t} (y_{u,t}|h_u)]
\end{split},
\end{equation}
% \yqc{Again, we cannot expect readers to read the reference. We need to explain the intuition behind.}
where the posterior distribution $q_{\phi_t}$ is estimated by parameterized neural networks $g_{\phi_t}$ which are trained by other users. Therefore, users do not update the estimator $g_{\phi_t}$ for private attributes. Specifically, we have:
\begin{equation}
\begin{split}
&\min_\theta \sum_{t \in \mathcal{T}_u} \max_{\phi_t} \mathbb{E}_{p(u, t)}[\log q_{\phi_t} (y_{u,t}|h_u)] \\
\approx & \max_\theta \sum_{t\in \mathcal{T}_u} CE(g_{\phi_t}(h_u), y_{u,t}) \text{.}
\end{split}
\end{equation}
% \yqc{Explain the above equation.}
As the labels $y_{u,t}$ are kept secret, the objective function is not practical in the training, we transform the objective function to unsupervised training. 
To maximize the cross entropy loss, we aim to raise the uncertainty of the estimated distribution:
\begin{equation}
\label{eqa:loss_pri}
\begin{split}
& \max_\theta \sum_{t\in \mathcal{T}_u} CE(g_{\phi_t}(h_u), y_{u,t})\\
\approx & \min_\theta \sum_{t \in \mathcal{T}_u} KL(g_{\phi_t}(f_\theta (u)), \hat{y}_{u,t}) \text{,}
\end{split}
\end{equation}
where $\hat{y}_{u,t} \sim\mathbb{U}$, which is the discrete uniform distribution.
    
\subsubsection{\textbf{Objective function}}
Equations (\ref{eqa:loss_rec}), (\ref{eqa:loss_non}), and (\ref{eqa:loss_pri}) are three sub-targets in UC-FedRec which correspond to three motivations respectively: recommendation performance, attribute distribution estimation, and private information elimination. For each user, we formulate two local training processes to meet the private representation training objective.

We solve the Equation (\ref{eqa:loss_non}) for attribute distribution estimator training. We sample a set of users $u$ from $\mathcal{U}_t$ and the corresponding attribute labels $y_{u,t}$, disentangle the training objective, and view the user representation $f_\theta (u)$ as the input. We have:
\begin{equation}
    \label{equ:filter}
    \begin{split}
        \phi_t = \arg \min_{\phi_t} \sum_{u \in \mathcal{U}_t} CE(g_{\phi_t}(h_u), y_{u,t}).
    \end{split}
\end{equation}

To solve the Equation (\ref{eqa:loss_rec}) and Equation (\ref{eqa:loss_pri}). We sample a set of users from $\mathcal{U}$ and their historical interaction data. Combining two objectives as joint learning, the private representation training objective for clients is as follows:
\begin{equation}
\label{equ:combine}
\begin{split}
        \theta, \psi = \arg \min_\theta (\beta \min_\psi \mathcal{L}& (s_\psi(f_\theta (u), f_\theta (i)), \mathbf{Y}) \\+ &(1-\beta) \sum_{t \in \mathcal{T}_u} KL(g_{\phi_t}(f_\theta (u)), \hat{y}_{u,t}) ) \text{,} 
\end{split}
\end{equation}
where $\beta \in [0,1]$ is the trade-off between recommendation utility and privacy protection. A larger $\beta$ indicates stronger privacy protection while a $smaller$ $\beta$ has a better recommendation accuracy. Note that the weight of each attribute's privacy loss can be adjusted. Here we assume that users weigh all the private attributes the same for simplicity.
% \begin{figure}[t]
%   \centering 
%   \includegraphics[width=\linewidth]{figure/distribute_training.pdf}
%   \caption{The training process of compositional privacy protection in clients. The blue arrows indicate computation, and the green arrows indicate backpropagation. For each user, there are two training objectives: (1) users training of the attribute distribution estimators (filters) on their own non-private attribute labels and (2) private representation training. } 
%   \label{Fig:description} 
% \end{figure}

% \subsubsection{\textbf{Differential privacy}\label{sec:LDP}}
%  Differential privacy is commonly used in federated learning to protect the transmission of gradients in training. We follow the technique proposed in \cite{qi2020privacy} to achieve the $\epsilon$-LDP guarantees. At each training round, the user $u$ generates gradients $\mathbf{g}_u$ for the recommendation network and non-private attribute filters, then perturb the clipped gradient using Laplace noise: 
%  \begin{equation}
%  \label{equ:LDP}
%      \bar{\mathbf{g}}_u = clip(\mathbf{g}_u, \delta) + Laplace(0, \lambda),
%  \end{equation}
% where $clip$ is 1-norm clip function. Users send new perturbed gradients $\bar{\mathbf{g}}_u$ to server. Using LDP technique, it is more difficult to recover the raw user interaction from the gradients. It is shown in \cite{qi2020privacy} that the upper bound of the privacy budget $\epsilon$ is $\frac{2\delta}{\lambda}$, which means that we can achieve stronger privacy guarantees by using a smaller clipping threshold $\delta$ or a larger noise strength $\lambda$.  

\subsection{Training Algorithm \label{sec:alg}} 
There are three types of neural network in UC-FedRec: recommendation representation network $f_\theta$, recommendation score function $s_\psi$ are learned for recommendation, and attribute distribution estimator $g_{\phi_t}$ is for sensitive attribute information protection. We adopt joint learning to train three networks. The UC-FedRec training process can be divided into two scenarios: distributed learning and central learning.

\subsubsection{\textbf{Distributed learning}}
To solve the learning objective proposed in Section \ref{sec:obj}, we make a complement to common FedRecs. Similar to a general federated learning system, the central server is responsible for coordinating the training process. For example, sampling a set of users participating in a training round, distributing parameters, aggregating and updating model weights \cite{mcmahan2017communication}. Compared to the FedRec used as the base model, the server takes on extra filters updating. The server needs to aggregate gradients from those non-private users for filters and distribute them to all the users for privacy-preserving learning. 

For clients, local training updates three types of neural networks. Clients perform several iterations to update non-private attribute filters weights $\phi_t, \forall t \in \mathcal{T}\backslash \mathcal{T}_u$, and compute the multi-task loss to update representation weights $\theta$ and score function weights $\psi$ in minibatch training. The filters eliminate specific user attribute information in representations. We iteratively update models' weights for local epoch $E$ times. Finally, the client adds Laplace noise to model weights and uploads the perturbed weights to the server. Similar as other federated learning system, the gradients are protected by LDP \cite{wu2021fedgnn, minto2021stronger}.

\subsubsection{\textbf{Central protection}\label{sec:cen}}
As users' privacy preferences change over time, UC-FedRec needs to provide protection promptly when users find an originally non-private attribute sensitive. To reduce the communication consumption, the server directly uses the learned filters to eliminate the sensitive information without federated learning. Assume that user $u$ requests additional attribute $t$ preservation, similar as the Equation (\ref{eqa:loss_pri}), we have the objective function: 
% \yqc{Again, $y_u,t$.}
\begin{equation}
    \min_{e_u} KL(g_{\phi_t}(u), \hat{y}_{u,t}).
\end{equation}
The propagation layer contains the collaborative signals, updating weights will inevitably influence others' recommendation accuracy, therefore, in the central protection, we choose to simply update user's embedding layer rather than the whole representation neural network to minimize the influence on the whole recommendation model. We iteratively update user embeddings until the difference reaches a threshold  ($|e_u'-e_u|  \leq T$).

\begin{table*}[]
\caption{Performance on different methods in terms of recommendation and privacy.}
\label{tab:main}
\begin{tabular}{cc|cccccccc}
\hline
\multicolumn{2}{c|}{\multirow{2}{*}{Methods}} & \multicolumn{5}{c}{MovieLens}                                                                                                & \multicolumn{3}{c}{Douban}                                            \\ \cline{3-10} 
\multicolumn{2}{c|}{}                         & \multicolumn{2}{c|}{Utility}                         & \multicolumn{3}{c|}{Privacy}                                          & \multicolumn{2}{c|}{Utility}                         & Privacy        \\ \hline
Protection      & Base Model                  & NDCG           & \multicolumn{1}{c|}{Recall}         & Gender         & Age            & \multicolumn{1}{c|}{Occupation}     & NDCG           & \multicolumn{1}{c|}{Recall}         & Location       \\ \hline
% -               & LDP-Rec                     & 0.694          & \multicolumn{1}{c|}{0.433}          & 0.685          & 0.296          & \multicolumn{1}{c|}{0.136}          & 0.634          & \multicolumn{1}{c|}{0.423}          & 0.233          \\
% -               & Fedfast-BPR                 & 0.72           & \multicolumn{1}{c|}{0.46}           & 0.745          & 0.353          & \multicolumn{1}{c|}{0.15}           & 0.684          & \multicolumn{1}{c|}{0.476}          & 0.255          \\ \hline
-               & \multirow{3}{*}{FedGNN}     & \textbf{0.843} & \multicolumn{1}{c|}{\textbf{0.544}} & 0.817          & 0.489          & \multicolumn{1}{c|}{0.171}          & \textbf{0.724} & \multicolumn{1}{c|}{\textbf{0.492}} & 0.265          \\
Early stopping      &                             & 0.794          & \multicolumn{1}{c|}{0.509}          & 0.787          & 0.364          & \multicolumn{1}{c|}{0.158}          & 0.692          & \multicolumn{1}{c|}{0.483}          & 0.249          \\
UC-FedRec       &                             & 0.783          & \multicolumn{1}{c|}{0.509}          & \textbf{0.631} & \textbf{0.283} & \multicolumn{1}{c|}{\textbf{0.128}} & 0.688          & \multicolumn{1}{c|}{0.482}          & \textbf{0.221} \\ \hline
-               & \multirow{3}{*}{FedNCF}     & \textbf{0.784}          & \multicolumn{1}{c|}{\textbf{0.496}}          & 0.755          & 0.358          & \multicolumn{1}{c|}{0.152}          & 0.696          & \multicolumn{1}{c|}{0.489}          & 0.263          \\
Early stopping      &                             & 0.742          & \multicolumn{1}{c|}{0.471}          & 0.653          & 0.284          & \multicolumn{1}{c|}{0.131}          & 0.686          & \multicolumn{1}{c|}{0.476}          & 0.245          \\
UC-FedRec       &                             & 0.734          & \multicolumn{1}{c|}{0.471}          & \textbf{0.602} & \textbf{0.212} & \multicolumn{1}{c|}{\textbf{0.117}} & 0.683          & \multicolumn{1}{c|}{0.477}          & \textbf{0.225} \\ \hline
\end{tabular}
\end{table*}

\section{Experiments \label{sec:exp}}

In this section, we conduct experiments to evaluate the performance of the proposed UC-FedRec. We aim to answer the following questions: \textbf{Q1:} Whether UC-FedRec can protect personalized privacy while maintaining high recommendation utility compared to its base FedRec model? \textbf{Q2:} How does UC-FedRec perform on different privacy preferences such as utility-privacy tradeoff, various private attributes, etc? \textbf{Q3:} Whether UC-FedRec can provide prompt protection when users' privacy preferences change? 
% \begin{itemize}
%     \item \textbf{Q1:} Whether UC-FedRec can protect personalized privacy while maintaining high recommendation utility compared to its base FedRec model?
%     \item \textbf{Q2:} How does UC-FedRec perform on different privacy preferences such as utility-privacy tradeoff, various private attributes, etc?
%     \item \textbf{Q3:} Whether UC-FedRec can provide prompt protection when users' privacy preferences change? 
% \end{itemize}

We first introduce experimental setup and metric in section \ref{sec:exp-setting}, then we evaluate UC-FedRec in sections \ref{sec:exp-Q1}, \ref{sec:exp-Q2}, and \ref{sec:exp-Q3} to answer three questions respectively. 

\subsection{Dataset and Experiment Setting \label{sec:exp-setting}}
\subsubsection{\textbf{Dataset}} We select two real-world recommendation datasets with users' side information: MovieLens \cite{harper2015movielens} and Douban \cite{ma2011recommender}. Both datasets are widely used in recommendation system evaluation. MovieLens contains 1,000,209 ratings by 6,040 users on 3,952 items. We treat users' gender, occupation and age as private attributes. Douban includes 647,263 interactions by 6,368 users with 22,347 items, location is treated as privacy. We transform the datasets into implicit data where each observed rating is treated as a positive instance and indicated by an interaction signal. Besides, we treat user attributes as sensitive information. As some users lack attribute information, we only retain users with all features so as for convenient privacy evaluation. For each user, we randomly select a set of features as the private attributes with probability $\alpha$ and the remaining as the non-private attributes. In such a way, we can simulate various compositional privacy preferences. In the evaluation, if there is no further statement, we set $\alpha = 0.3$.

\subsubsection{\textbf{Hyperparameter setting}}
 In the experiment, if there is no further statement, we use the following implementation settings. We use FedGNN \cite{wu2021fedgnn} and FedNCF \cite{perifanis2022federated} as our base FedRecs and use dot product as the scoring function. The dimension of user and item embeddings and their hidden representations learned by FedRecs are 128. We follow the technique proposed in \cite{qi2020privacy} to achieve the $1$-LDP guarantees to protect the gradient transmission. We use SGD as the optimization algorithm with 0.1 learning rate. We use BPR loss to train the recommendation model. For the privacy-preserving part, we use $2$-layer perceptrons (MLP) as attribute information filters and information leakage evaluation using SGD optimization with 0.01 learning rate. We set the privacy-utility tradeoff $\beta=0.5$.
\begin{figure*}

\centering
\subfigure[Evaluation on gender and HR@10]{\includegraphics[width=0.3\textwidth]{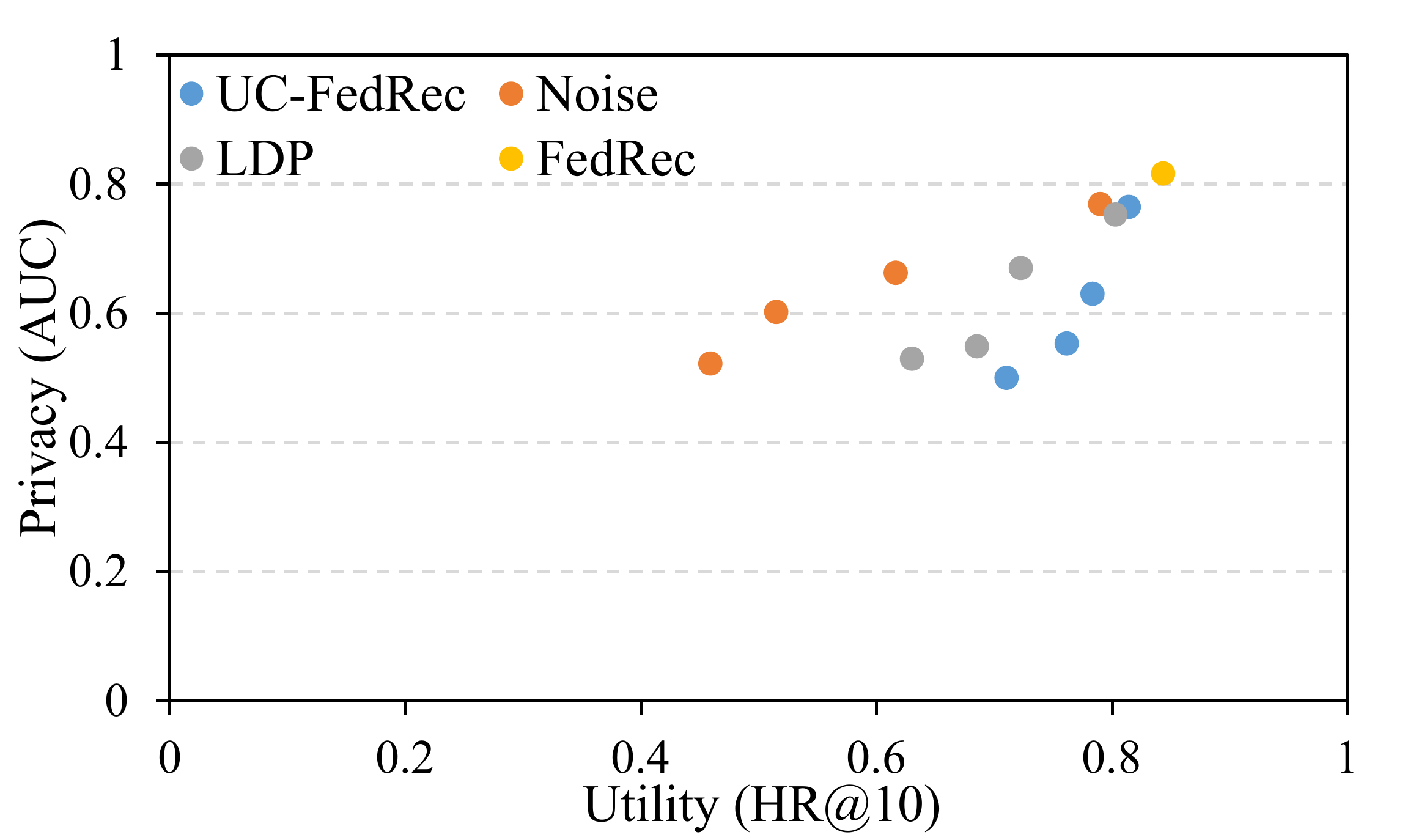}
\label{subfig:genderhr}}\hfil
\subfigure[Evaluation on age and HR@10]{\includegraphics[width=0.3\textwidth]{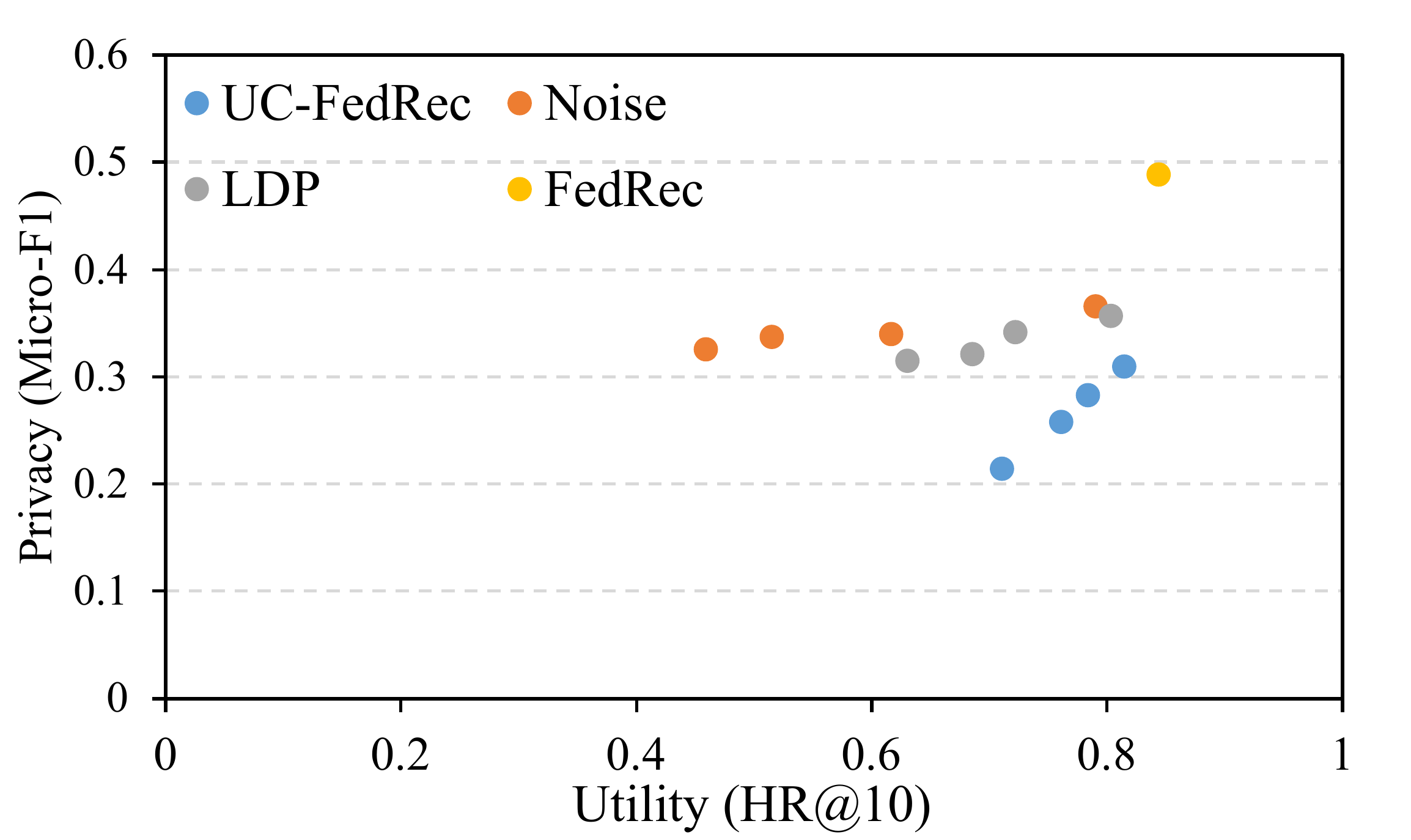}}\hfil 
\subfigure[Evaluation on occupation and HR@10]{\includegraphics[width=0.3\textwidth]{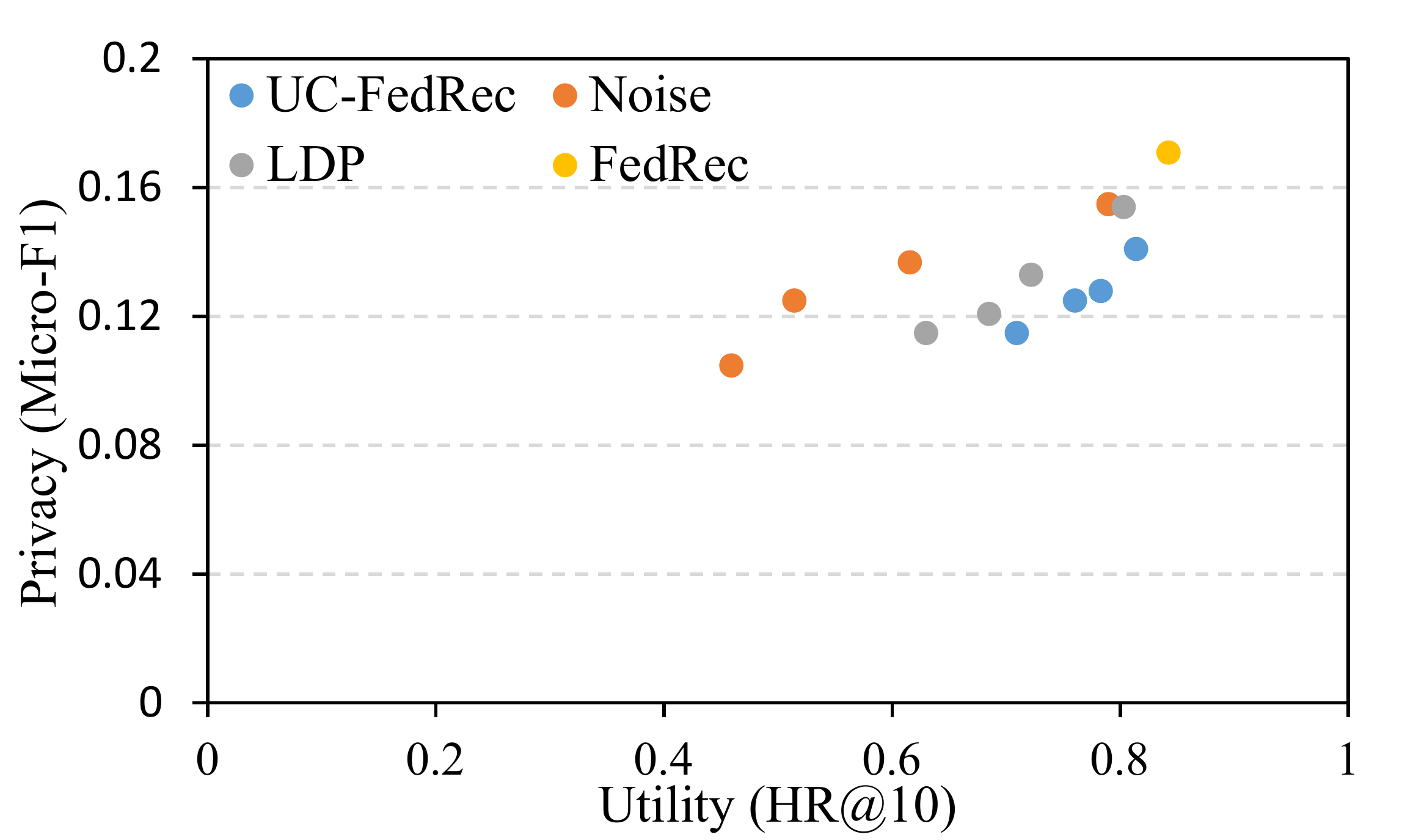}} 

\subfigure[Evaluation on gender and NDCG@10]{\includegraphics[width=0.3\textwidth]{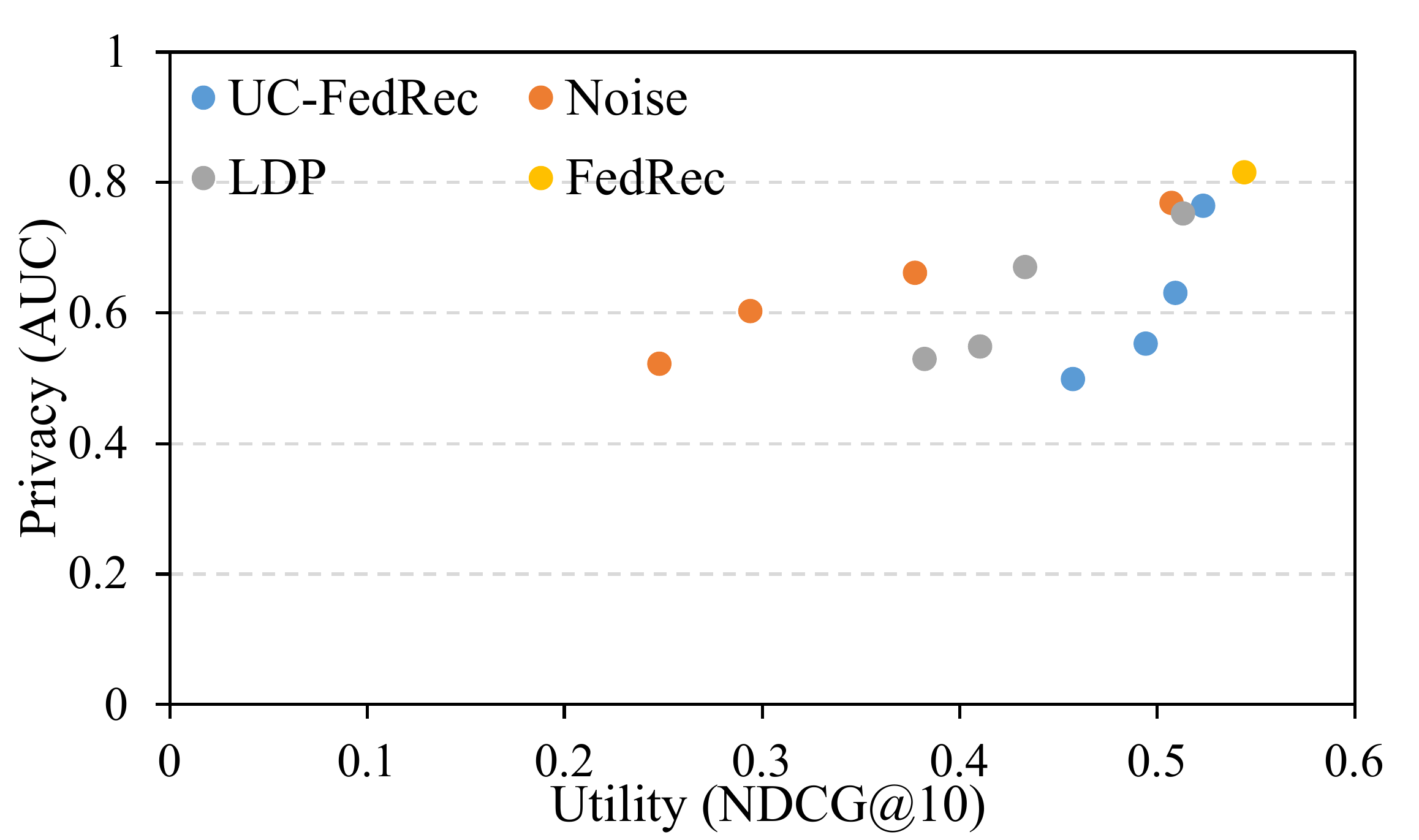}\label{subfig:genderndcg}}\hfil   
\subfigure[Evaluation on age and NDCG@10]{\includegraphics[width=0.3\textwidth]{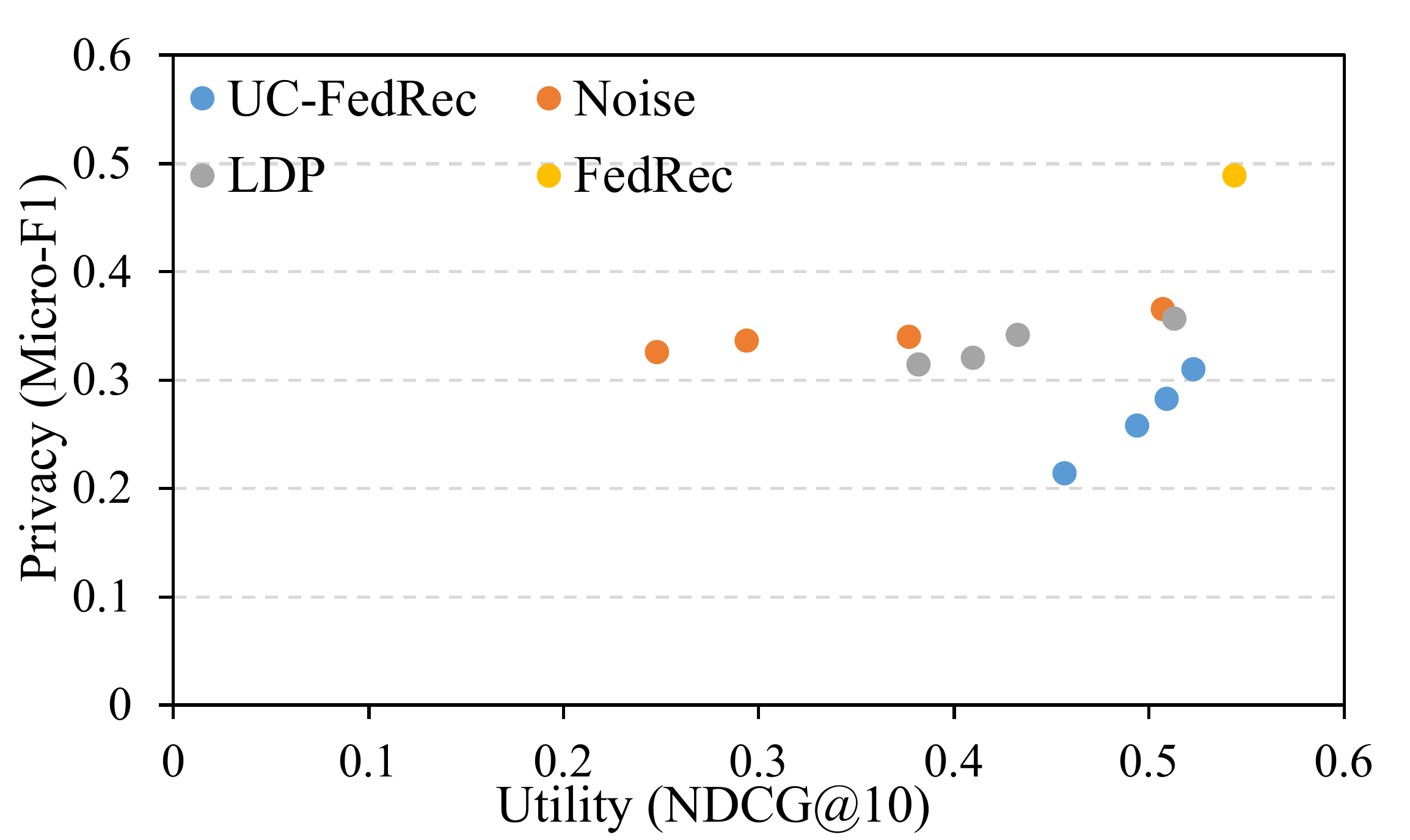}}\hfil
\subfigure[Evaluation on occupation and NDCG@10]{\includegraphics[width=0.3\textwidth]{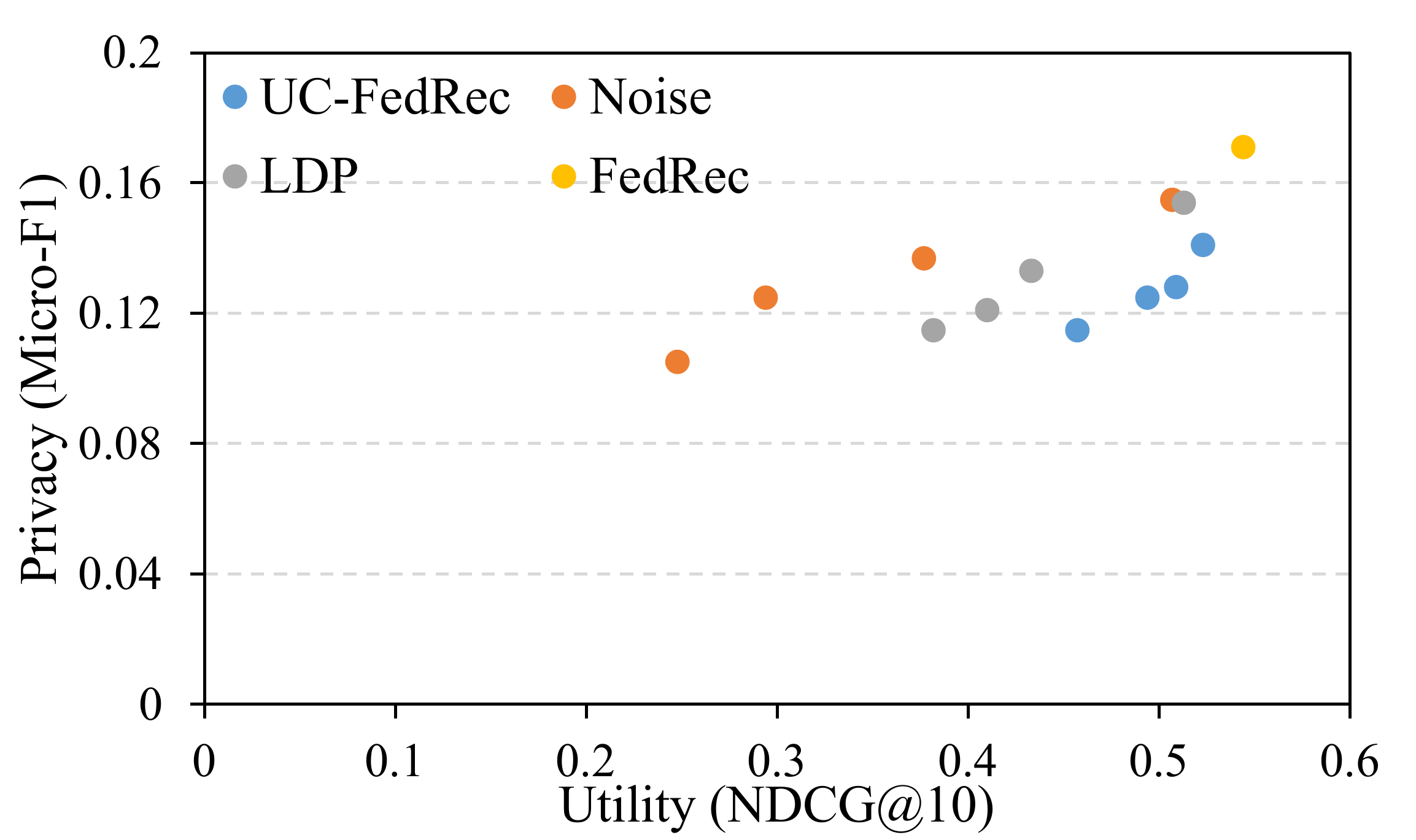}}
\caption{Utility-privacy tradeoff comparison of the framework with baselines on MovieLens.}
\label{Fig:baseline}
\end{figure*}
\subsubsection{\textbf{Performance Evaluation}}
To evaluate recommendation utility, we adopt the \textit{leave-one-out} strategy which has been widely used in literatures \cite{deshpande2004item}. We take out their latest interaction for the test set and use the remaining data for training. Since ranking all the items is time-consuming, we follow the common strategy \cite{elkahky2015multi, he2017neural} that randomly samples 50 items that are not interacted with by the user and ranks the test item among the 50 items. The performance of a ranked list is adjusted by \textit{Hit Ratio} (HR) \cite{deshpande2004item} and \textit{Normalized Discounted Cumulative Gain} (NDCG). We truncate the ranked list at 10 for both metrics. We calculate both metrics for each test user and report the average score. For privacy protection evaluation, we adopt attribute classifiers to evaluate the information leakage in the representations. We use neural networks to infer users' attributes. Binary attributes are evaluated using \textit{Area under the ROC Curve} (AUC) and multi-class attributes are evaluated using \textit{micro-F1}. Our goal is to prevent features from inference attacks, therefore lower scores indicate better privacy protection.

\subsection{\textbf{Problem Evaluation}}
We evaluate the attribute inference risks in several FedRecs and compare UC-FedRec's utility performance and private attribute protection ability to them. We select FedGNN \cite{wu2021fedgnn}, a graph-based FedRec on user-item graph expansion and private sharing, and FedNCF \cite{perifanis2022federated}, a federated learning extension of neural collaborative filtering as our baselines.
% \begin{itemize}
%     \item FedGNN \cite{wu2021fedgnn}, a graph-based FedRec on user-item graph expansion and private sharing.
%     \item FedNCF \cite{perifanis2022federated}, a federated learning extension of neural collaborative filtering.
%     % \item LDP-Rec \cite{minto2021stronger}, a FedRec based on $\epsilon-$LDP privacy protection.
%     % \item FedFast \cite{muhammad2020fedfast}, a FedRec based on active aggregation method and client sample strategy.
% \end{itemize}
We implement these FedRecs basically following the original paper while setting the same embedding dimension for fair comparison. We apply UC-FedRec framework to both base models to evaluate the effectiveness. Besides, we also report the performance of FedGNN and FedNCF with early stopping to maintain similar recommendation performance (recall) for convenient comparison. The results are summarized in Table \ref{tab:main}. 
% Firstly, we observe that all FedRecs face the problem of private attribute attack, a FedRec with better recommendation performance also faces a more serious information leakage problem. This is probably because those high-quality representations improve the recommendation performance as well as contain personal information. 
Compared to the base model, UC-FedRec can efficiently protect private information while sacrificing acceptable recommendation accuracy. With similar recommendation utility, models with UC-FedRec better protect private attributes than early-stopping. Besides, UC-FedRec provides privacy protection to both FedGNN and FedNCF, and its performance is related to the base model. UC-FedRec with FedGNN as its base model can perform better in recommendation utility while facing relatively severe privacy leakage than FedNCF.

 \subsection{\textbf{Model Effectiveness} \label{sec:exp-Q1}}
 We also validate our method's effectiveness in privacy preservation. To the best of our knowledge, there are no techniques that can provide personalized and compositional privacy protection in federated learning, therefore, we only select 2 types of data privacy-preserving baselines \cite{liu2019privacy} and compare our framework's overall protection performance on all attributes with them for a fair comparison. A detailed introduction of these baselines is listed below:  
 
 \textbf{Noise perturbation:} We train the base FedRec and apply Gaussian noise $\mathcal{N}(0, \sigma^2)$ to trained private user embeddings in the server. The injected noise can provide strong privacy guarantees that avoid the model from privacy leakage. The tradeoff can be influenced by noise strength $\sigma$.  
 
\textbf{LDP:} LDP is widely used to protect privacy, user injects Laplace noise to the gradients on the client side, and the privacy budget can be adjusted by the noise strength $\lambda$, therefore, we inject stronger noise to private user's gradients to provide better protection.  

We compare the utility privacy tradeoff in our framework with the other two baselines. In our experiments, we provide three feature protection choices in MovieLens dataset, which are gender, age, and occupation. As the two baselines cannot provide personalized information preservation, we assume that there are 30\% users who want to hide all their attributes and other users do not mind revealing some personal attribute information to have the best recommendation experience for a fair comparison. For our framework, we set different $\beta$ for various utility-privacy balances. Here we set $\beta = \{0.3, 0.5, 0.7, 0.8\}$, respectively. For the noise perturbation baseline, we add different strength Gaussian noise to learned private user embeddings. Here we set $\sigma=\{1,3,5, 7\}$, respectively. In the LDP method, a smaller privacy budget provides stronger privacy guarantees for users. We differently set private users' smaller privacy budgets to protect their privacy. Here we set $\epsilon = \{0.1, 0.3, 0.5, 0.7\}$, respectively. We evaluate and compare private users' recommendation utility and privacy protection.
\begin{figure}[t]
    \centering
    \subfigure[Evaluation on MovieLens.]{\includegraphics[width=0.49\linewidth]{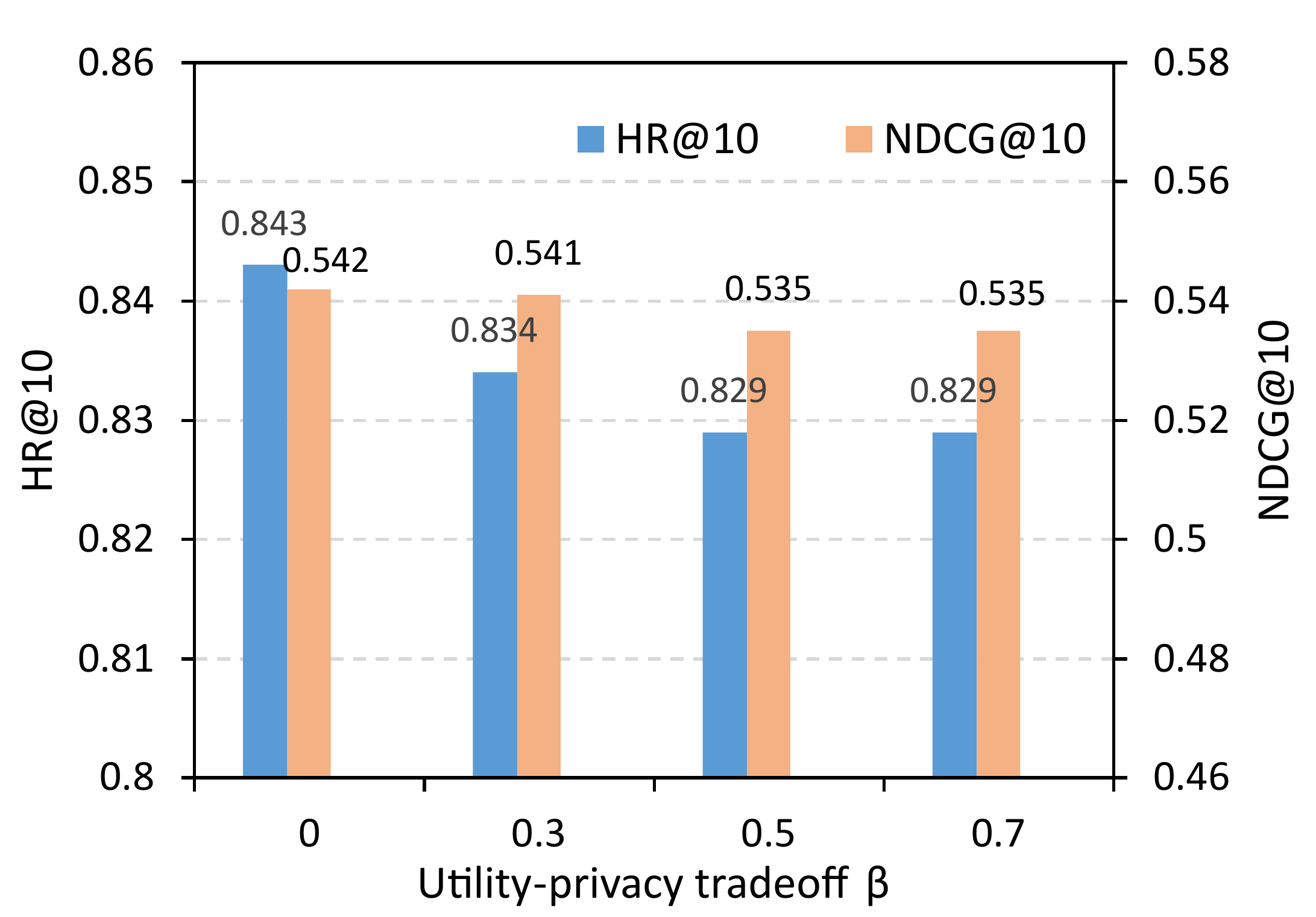}}
    \subfigure[Evaluation on Douban.]{\includegraphics[width=0.49\linewidth]{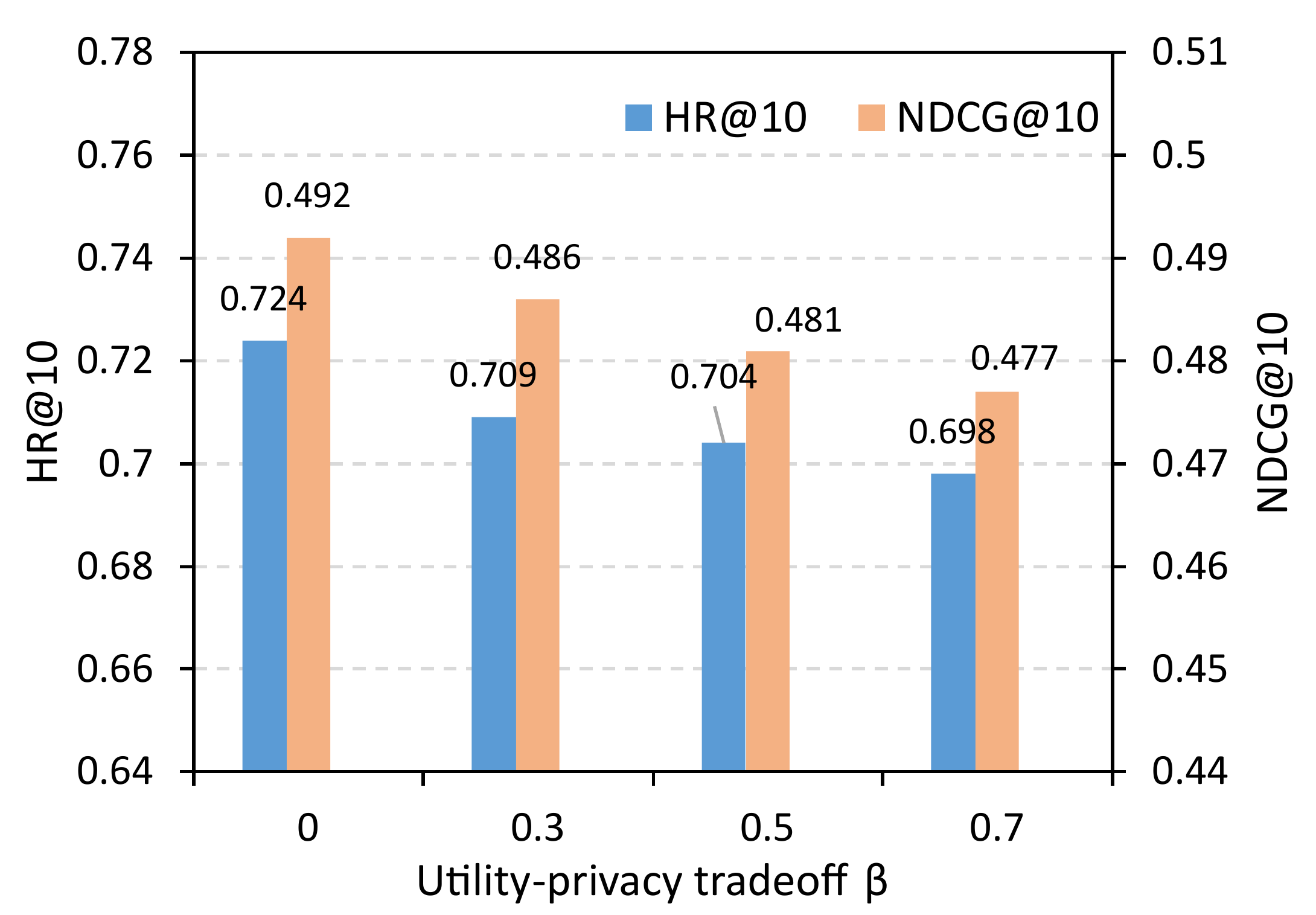}}
    \caption{Utility evaluation on non-private users.}
    \label{Fig:non-private}
\vspace{-0.4cm}
\end{figure}
As shown in Figure \ref{Fig:baseline}, it can be observed that privacy protection leads to a degradation in utility in all the techniques. However, our framework has a better utility-privacy tradeoff compared to the other two baselines. Utilizing the different user privacy preferences, the framework can eliminate the privacy information more accurately. Taking gender protection as an example, in Figure \ref{subfig:genderhr} and \ref{subfig:genderndcg}, the AUC of Gender inference attack on the base FedRec is $0.817$ and decreases to $0.5$ when the utility-privacy tradeoff $\beta$ increases to $0.8$. When our framework can eliminate all the private users' gender information ($\text{AUC} =0.5$), it can still remain $0.75$ HR and $0.457$ NDCG which is $89.0\%$ and $84.0\%$ of the base model performance respectively. While LDP and noisy perturbation can only get about $72\%$ and $50\%$ original recommendation performance. 

\subsection{Privacy Influence \label{sec:exp-Q2}}

% \begin{figure}[t]
%   \centering
%   \includegraphics[width=\linewidth]{figure/non-private_new.png}
%   \caption{The recommendation performance evaluation on non-private users.}
%   \label{Fig:non-private}
% \end{figure}

\subsubsection{\textbf{Influence on non-private users.}}
Users have different privacy preferences and some users care more about their recommendation experience rather than personal privacy, therefore the recommendation utility for them cannot be largely influenced. We conduct experiments to evaluate our framework's influence on non-private users. For every user, we randomly sample private attributes with a probability of $30\%$, and evaluate the recommendation performance for those users with no private attributes. 

Figure \ref{Fig:non-private} shows the recommendation performance for non-private users in MovieLens and Douban datasets. It indicates that our framework can well retain recommendation performance for non-private users. These users do not need to pay much utility cost for unnecessary privacy. For example, when $\beta=0.7$, the HR and NDCG can still reach $0.698$ and $0.477$ which are $96.4\%$ and $97.0\%$ for non-private users compared to the base model in Douban dataset respectively. The recommendation performance still degrades due to the framework will eliminate the sensitive information in the whole system including collaborative signals. 

\subsubsection{\textbf{Personalized privacy preference.}} 
\begin{figure}[t]
    \centering
    \subfigure[Gender Protection.]{\includegraphics[width=0.48\linewidth]{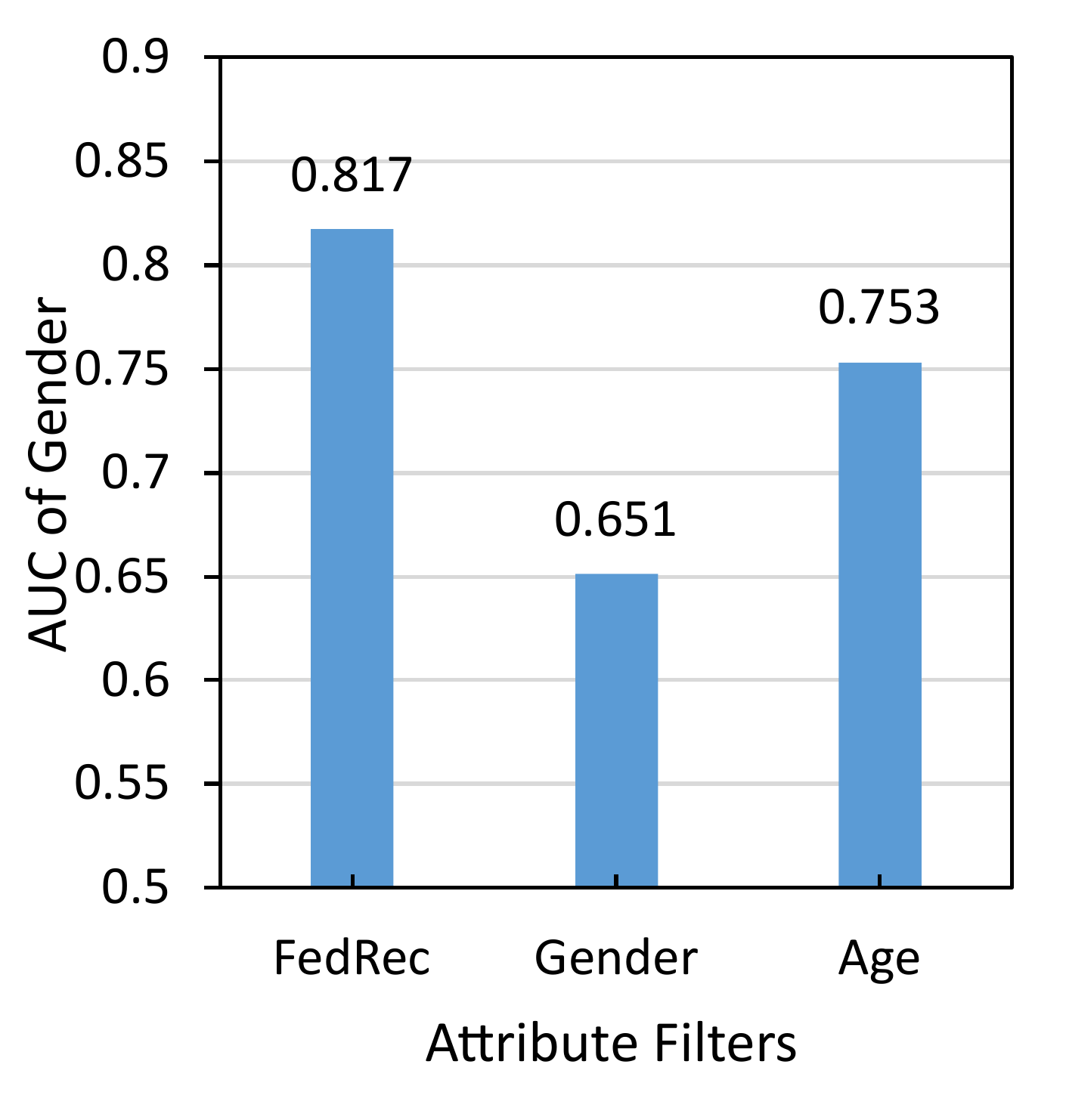}}
    \subfigure[Age Protection.]{\includegraphics[width=0.48\linewidth]{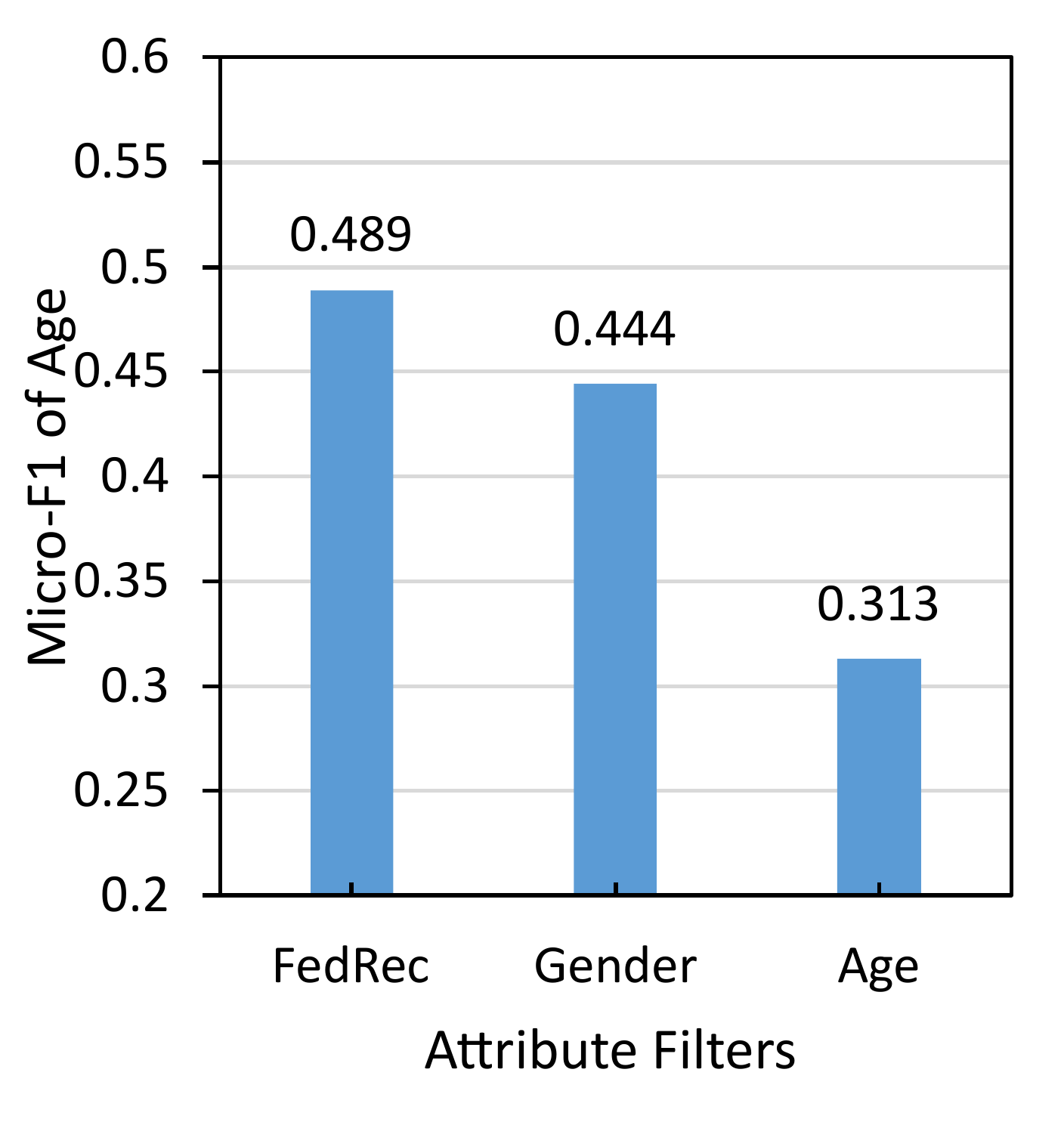}}
    \caption{The personalized protection of private attributes.}
    \label{Fig:single}
\vspace{-0.4cm}
\end{figure}

Besides users' different privacy and utility preferences, they usually have various privacy preferences. For example, some users may find gender is more sensitive compared to age while others do not. We conduct experiments on single attribute protection to validate personalized privacy protection. In MovieLens, we randomly assign gender or age privacy protection for every user and infer their personal attributes. As shown in Figure \ref{Fig:single}, both gender and age feature filters can provide privacy protection for either gender and age compared to the base model. For example, the gender filter eliminates some age information so that the attacker's inference micro-F1 drops from $0.489$ to $0.444$. However, the age filter can protect users' age information more accurately where the attacker can only reach $0.313$ mirco-F1, which means that our specialized filters can provide more precise protection to meet users' different personalized privacy preferences.

\subsubsection{\textbf{Private attribute amounts.}}
\begin{figure}[t]
    \centering
    \includegraphics[width=0.8\linewidth]{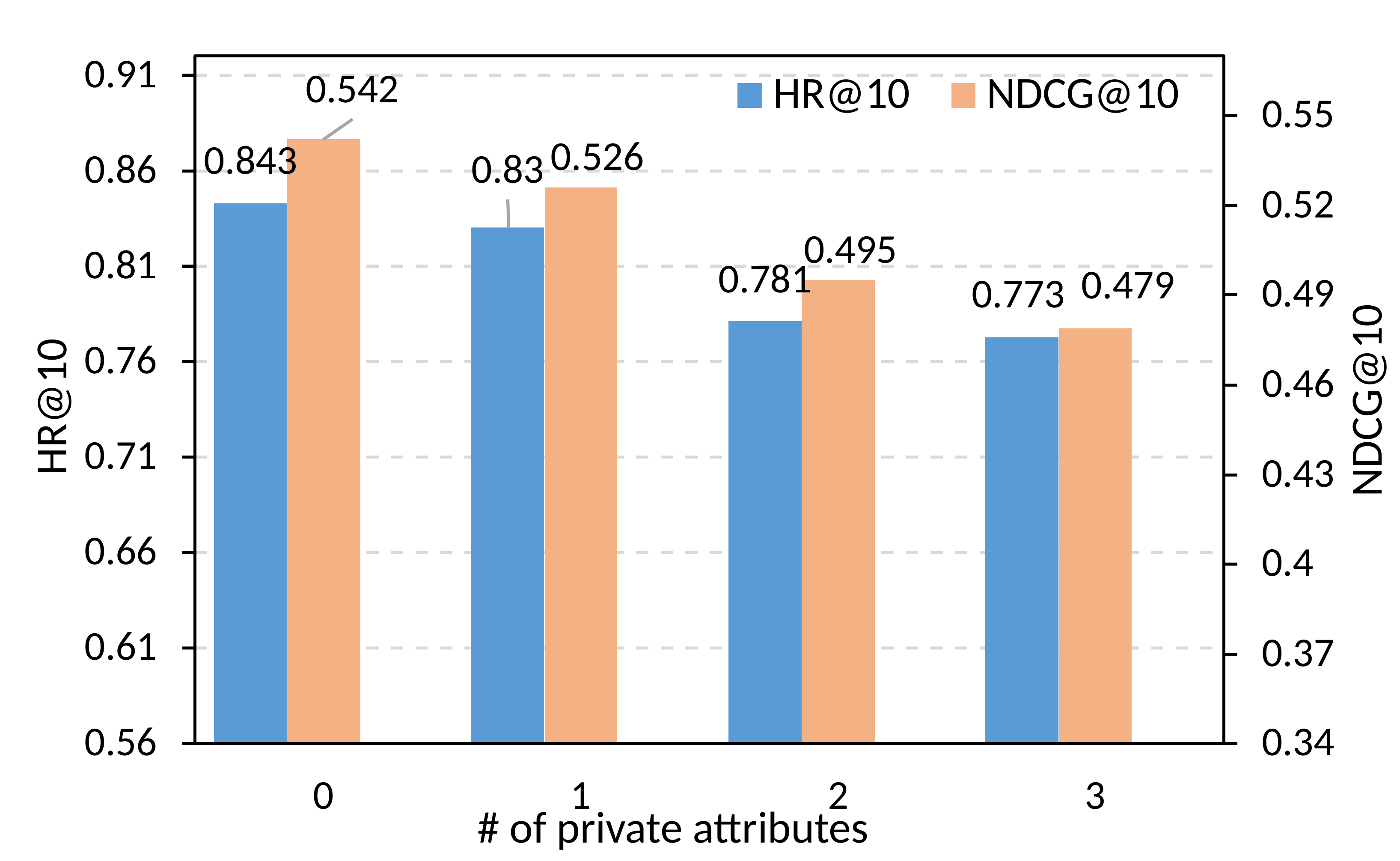}
    \caption{Recommendation utility evaluation on the number of private attributes.}
    \label{Fig:num_attribute}
\vspace{-0.4cm}
\end{figure}

In this part, we evaluate the impact of the number of private attributes. We divide users into different groups according to their private attribute amount, then evaluate the recommendation performance of each group.
% \begin{figure}[t]
%   \centering
%   \includegraphics[width=\linewidth]{figure/num_attribute_new.png}
%   \caption{Recommendation performance evaluation on the number of private attributes.}
%   \label{Fig:num_attribute}
% \end{figure}
\begin{figure}[t]
    \centering
    \subfigure[HR@10 evaluation.]{\includegraphics[width=0.49\linewidth]{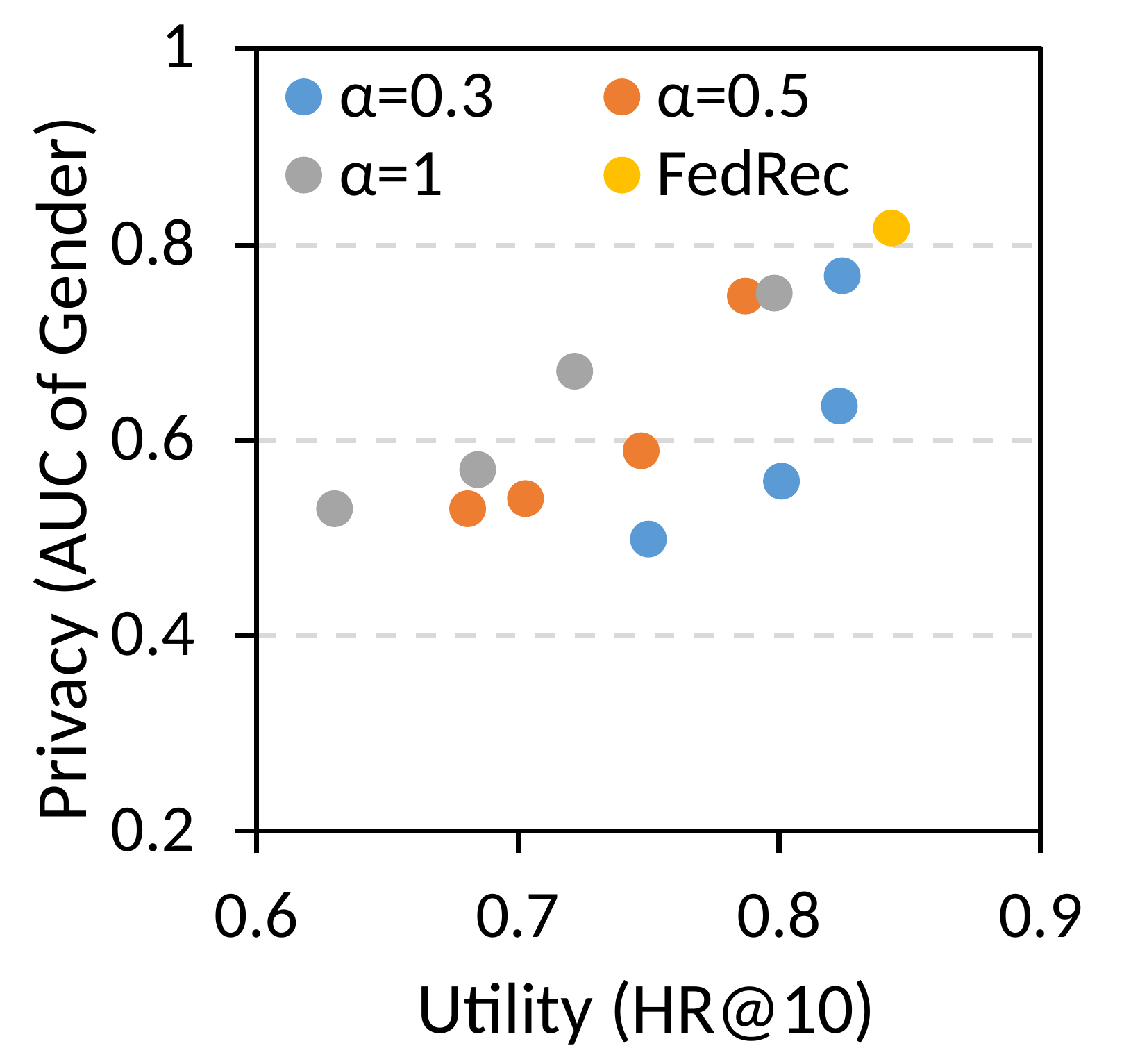}}
    \subfigure[NDCG@10 evaluation.]{\includegraphics[width=0.49\linewidth]{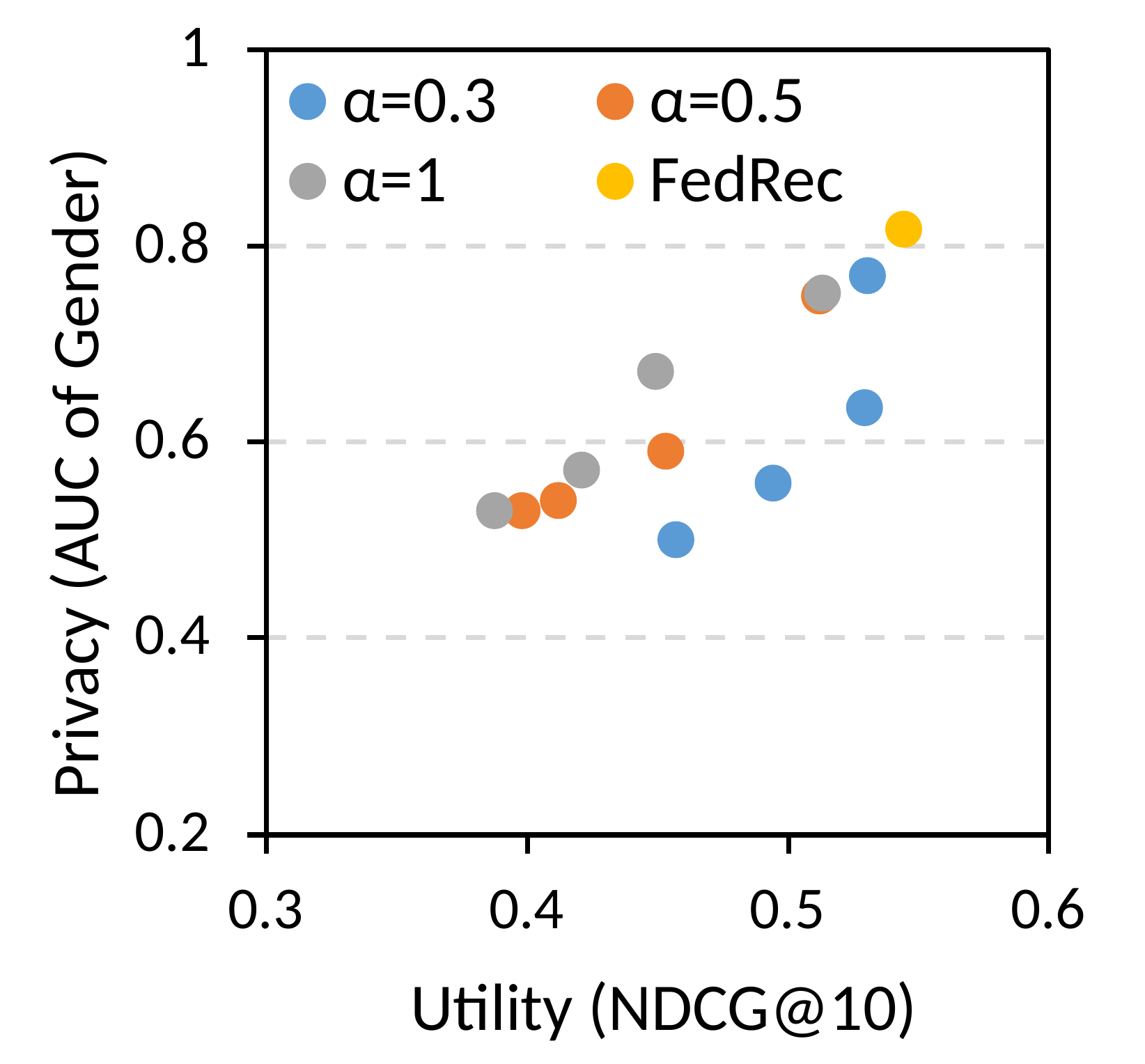}}
    \caption{Comparison of different private user ratios.}
    \label{Fig:pri_ratio}
\vspace{-0.4cm}
\end{figure}
As shown in Figure \ref{Fig:num_attribute}, it can be observed that the recommendation utility degrades when the private attributes increase, which means that users pay more recommendation costs when they need to protect more attributes. However, it still remains high recommendation accuracy when all the attributes are protected. For example in MovieLens, users can have $0.773$ HR and $0.479$ NDCG which are $91.7\%$ and $88.3\%$ compared to non-private users respectively, it indicates that UC-FedRec can retain most of the collaborative filtering information and satisfy users' privacy needs with acceptable cost.

\subsubsection{\textbf{Private user ratio.}}
As our framework utilizes users' different privacy preferences and needs users to share part of their non-private attributes to train the attribute distribution estimators. The ratio of private attributes for the whole recommendation system can influence the estimators' accuracy, thereby influencing the protection effectiveness. Therefore, we evaluate the framework on different private attribute ratios $\alpha$. We choose privacy ratio $\alpha=\{0.3, 0.5, 1\}$ respectively and evaluate the privacy protection and recommendation utility under different tradeoffs.

% \begin{figure}[t]
%   \centering
%   \includegraphics[width=\linewidth]{figure/pri_ratio_new.png}
%   \caption{Comparison of different private attribute ratio.}
%   \label{Fig:pri_ratio}
% \end{figure}

Figure \ref{Fig:pri_ratio} shows that UC-FedRec can protect users' privacy under all the private users' ratios. However, if there are more users willing to share their own attributes, the framework can provide a better utility-privacy tradeoff balance for users, because the attribute distribution estimators are well-trained and the private information is eliminated precisely. When the private attribute ratio increases, the framework has to pay more recommendation utility to get the same protective effect. For example, though the attack AUC is similar ($\text{AUC} = 0.56$), the recommendation utility can retain $\text{HR} = 0.75$ when $\alpha=0.3$ while $\text{HR} = 0.63$ when $\alpha = 1$. This is because of the performance drops in sensitive attribute estimators. Besides, even if there are no user-sharing attributes ($\alpha=1$), the framework can still provide protection because it adds random noise to the gradients and performs like an LDP protection under this circumstance. 

\subsection{\textbf{Central Protection} \label{sec:exp-Q3}}
In the framework, privacy protection can be done not only during the training stage but also for the trained model. As introduced in Section \ref{sec:cen}, when users' privacy preferences change, the framework can directly use trained filters to provide protection promptly.
% \yqc{Not clear here. Please refer to the corresponding section when introducing the approach.} 
After the model is trained, we randomly choose $20\%$ users who do not care about specified feature protection and apply central protection to protect the feature for these users. We set the threshold $T=100$ and iteratively update the embeddings of these users until the difference reaches the threshold. Table \ref{tab:central} shows the central protection performance on the age attribute in MovieLens and Occupation attribute in Douban. The utility and privacy evaluation comparison indicates that the framework can remove sensitive information on a central server with acceptable recommendation utility costs. 
% When users' privacy preferences are changed, we do not need to re-train the whole model, applying private attribute filters at the central server can provide prompt preservation to those users.

\begin{table}[t]
\caption{Results on central protection.}
\label{tab:central}
\begin{tabular}{cc|cc|c}
\hline
                                                                              & \multirow{2}{*}{Protection} & \multicolumn{2}{c|}{Utility} & Privacy  \\
                                                                              &                             & HR            & NDCG         & Micro-F1 \\ \hline
\multirow{2}{*}{\begin{tabular}[c]{@{}c@{}}MovieLens \\ (Age)\end{tabular}}   & Non-private                 & 0.829         & 0.535        & 0.387    \\
                                                                              & Private                     & 0.768         & 0.502        & 0.291    \\ \hline
\multirow{2}{*}{\begin{tabular}[c]{@{}c@{}}Douban \\ (Location)\end{tabular}} & Non-private                 & 0.721         & 0.492        & 0.304    \\
                                                                              & Private                     & 0.698         & 0.473        & 0.265    \\ \hline
\end{tabular}
\end{table}

\begin{figure}[t]
  \centering
  \includegraphics[width=0.9\linewidth]{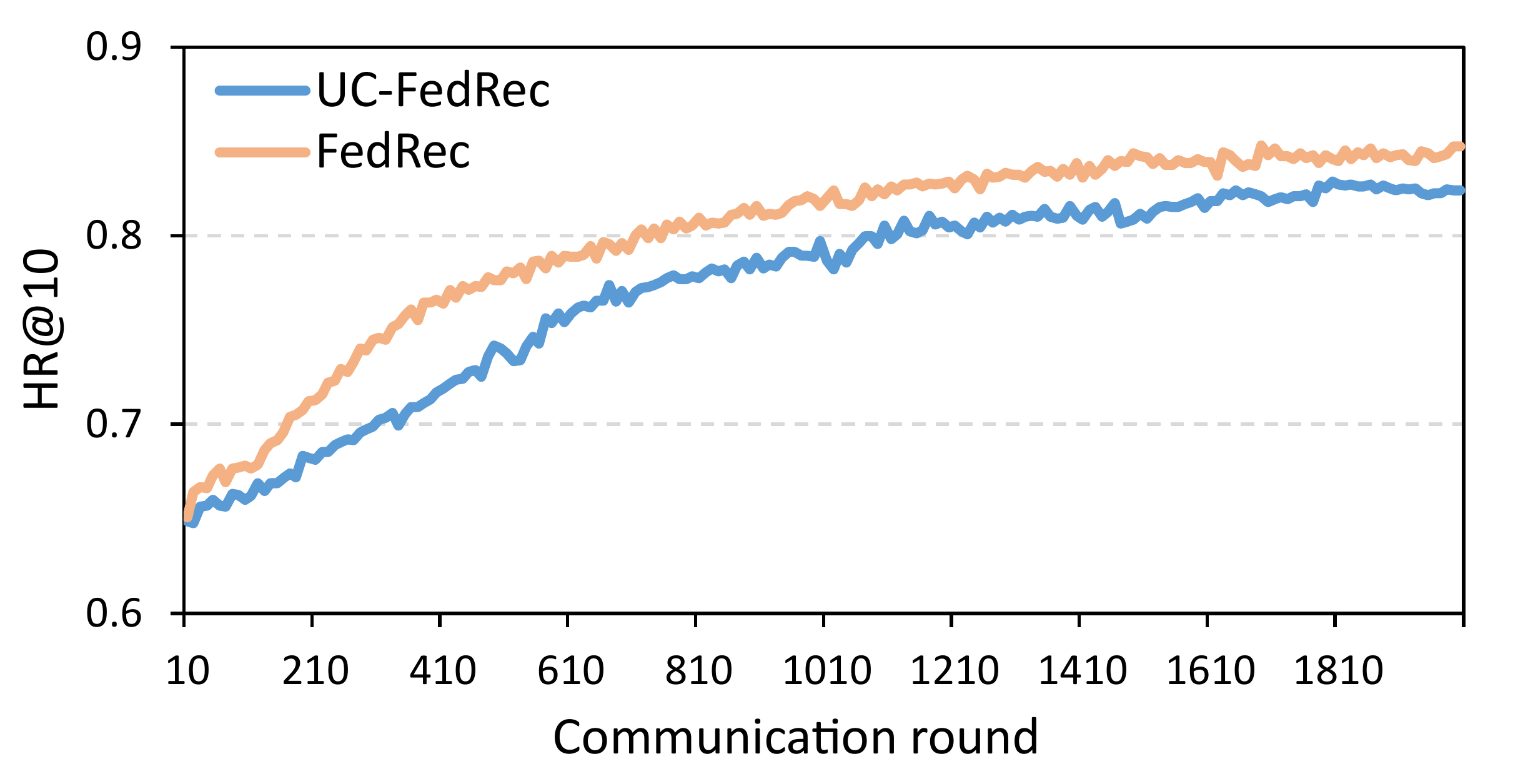}
  \caption{Convergence speed on MovieLens.}
  \label{Fig:convergence}
  \vspace{-0.4cm}
\end{figure}

\subsection{Convergence Rate \label{sec:conv}}
In this section, we evaluate whether the framework would influence the convergence rate. Figure \ref{Fig:convergence} shows the framework's recommendation performance in different communication rounds compared to the base model in MovieLens. As shown in the figure, the framework slightly influences the convergence speed, including slower convergence speed and less stable training. However, the difference mainly happens at the beginning of training. It is mainly because the attribute distribution estimators need several rounds to converge. The total communication rounds for the FedRec and our UC-FedRec are similar. 
% \yqc{It's better to name your framework.}

\section{Conclusion \label{sec:conclude}}

In this paper, we present a personalized privacy protection framework for FedRec which handles the problem that traditional FedRec suffers inference attacks. Our framework can provide different user-level attribute preservation according to users’ various privacy preferences. Experiments on two real-world datasets demonstrate that our framework outperforms the baselines in recommendation utility and privacy protection tradeoff. Our framework can provide flexible and effective privacy protection which does not pay much recommendation accuracy cost. In addition, the framework's convergence speed is similar to the base FedRec. In the future, we plan to extend our model to non-adversarial techniques and unsupervised learning scenarios to protect unseen attributes. 

\section*{ACKNOWLEDGMENTS}
The authors of this paper were supported by the NSFC Fund (U20B2053) from the NSFC of China, the RIF (R6020-19 and R6021-20) and the GRF (16211520 and 16205322) from RGC of Hong Kong. We also thank the support from Webank and the UGC Research Matching Grants (RMGS20EG01-D, RMGS20CR11, RMGS20CR12, RMGS20EG19, RMGS20EG21, RMGS23CR05, RMGS23EG08).

\bibliographystyle{ACM-Reference-Format}
\balance
\bibliography{sample-base}
\section*{Ethical Considerations}
UC-FedRec frameworks can effectively protect users' privacy from attribute inference attacks if the technology is being used as intended. However, we do not consider the situation that the FedRec server is malicious where some shared information may face greater risks. The regulations of service providers are needed and technical solutions remain future works.  

 

\end{document}